\documentclass[11pt]{article}

\usepackage[utf8]{inputenc}
\usepackage{amsmath}
\usepackage{bm}
\usepackage{cancel}  
\usepackage{amssymb}
\usepackage{amsthm}
\usepackage[titletoc,toc,title]{appendix}
\usepackage{array}
\usepackage{authblk}
\usepackage{bbm}
\usepackage{caption}
\usepackage{dsfont}
\usepackage{setspace}
\usepackage{enumitem}
\usepackage[margin=1in]{geometry}
\usepackage{graphicx}
\usepackage{hyperref}
\usepackage{layout}
\usepackage{mathtools}
\usepackage{multirow}
\usepackage{afterpage}
\usepackage[table,xcdraw]{xcolor}
\usepackage{url}
\usepackage[
backend=biber, natbib,
citestyle = authoryear,
style = authoryear,
uniquelist=false,
uniquename = false,
sorting = nyt,
maxbibnames=99,
maxcitenames=2,
]{biblatex}
\renewbibmacro{in:}{}
\usepackage{leftidx}
\usepackage{xcolor}
\usepackage{subcaption}
\usepackage{tikz}
\usepackage[ruled,linesnumbered]{algorithm2e}
\usepackage{booktabs}
\usepackage{lscape}
\usepackage{mathrsfs}

\usetikzlibrary{shapes, shadows, arrows, positioning, decorations.markings, decorations.pathreplacing, matrix, calc}

\newcommand*\circled[1]{\tikz[baseline=(char.base)]{
\node[shape=circle,draw,inner sep=1pt] (char) {#1};}}
\newcommand{\Depth}{4}
\newcommand{\Height}{4}
\newcommand{\Width}{4}
\usetikzlibrary{arrows.meta}

\makeatletter
\providecommand\phantomcaption{\caption@refstepcounter\@captype}
\makeatother

\tikzset{
->-/.style={decoration={
markings,\theoremstyle{plain}
\newlist{casenv}{enumerate}{4}
\setlist[casenv]{leftmargin=*,align=left,widest={iiii}}
\setlist[casenv,1]{label={{\itshape\ \casename} \arabic*.},ref=\arabic*}
\setlist[casenv,2]{label={{\itshape\ \casename} \roman*.},ref=\roman*}
\setlist[casenv,3]{label={{\itshape\ \casename\ \alph*.}},ref=\alph*}
\setlist[casenv,4]{label={{\itshape\ \casename} \arabic*.},ref=\arabic*}
mark=at position .5 with {\arrow{>}}},postaction={decorate}},
-<-/.style={decoration={
markings,
mark=at position .5 with {\arrow{<}}},postaction={decorate}},
}

\title{\LARGE\bf Cyber Risk Assessment for Capital Management}
\author[$\dagger$]{Wing Fung Chong}
\author[$\star$]{Runhuan Feng}
\author[$\ddagger$]{Hins Hu}
\author[$\diamond$]{Linfeng Zhang\thanks{Corresponding author}}
\affil[$\dagger$]{Maxwell Institute for Mathematical Sciences and Department of Actuarial Mathematics and Statistics, Heriot-Watt University. Email: alfred.chong@hw.ac.uk.}
\affil[$\star$]{Department of Finance, Tsinghua University. Email: fengrh@tsinghua.edu.cn.}
\affil[$\ddagger$]{Systems Engineering, Cornell University. Email: zh223@cornell.edu.}
\affil[$\diamond$]{Department of Mathematics, The Ohio State University. Email: zhang.14673@osu.edu.}
\begin{document}


\theoremstyle{definition}
\newtheorem{theorem}{Theorem}[section]
\newtheorem{corollary}[theorem]{Corollary}
\newtheorem{lemma}[theorem]{Lemma}
\newtheorem{proposition}[theorem]{Proposition}

\newtheorem{definition}{Definition}[section]
\newtheorem{problem}{Problem}
\newtheorem{remark}{Remark}[section]
\newtheorem{example}{Example}[section]
\setcounter{section}{0}
\date{}
\maketitle

\begin{abstract}

This paper introduces a two-pillar cyber risk management framework to address the pervasive challenges in managing cyber risk. The first pillar, cyber risk assessment, combines insurance frequency-severity models with cybersecurity cascade models to capture the unique nature of cyber risk. The second pillar, cyber capital management, facilitates informed allocation of capital for a balanced cyber risk management strategy, including cybersecurity investments, insurance coverage, and reserves. A case study, based on historical cyber incident data and realistic assumptions, demonstrates the necessity of comprehensive cost-benefit analysis for budget-constrained companies with competing objectives in cyber risk management. In addition, sensitivity analysis highlights the dependence of the optimal strategy on factors such as the price of cybersecurity controls and their effectiveness. The framework's implementation across a diverse range of companies yields general insights on cyber risk management.



\vspace{3mm}

\end{abstract}

\section{Introduction} \label{sec:intro}
As modern businesses and public-sector entities have become greatly reliant on information technology (IT) to boost the efficiency of workflows and stay connected with the world, potential cyber incidents, such as data breaches, pose great dangers to organizations' daily operations. Therefore, managing cyber risk should be a critical component in enterprise risk management practices.

\subsection{Difficulties in cyber risk management}
A successful cyber risk management plan should include both a risk assessment component and a capital allocation component, as noted in \citet{NIST2018} regarding how to establish or improve a cybersecurity program and how to make buying decisions. In practice, neither of the two tasks is trivial.

Regarding the risk assessment aspect, as an emerging type of risk, cyber risk is more difficult to assess than many other traditional risks for reasons including {\color{black} 1) lack of quantitative tools for comprehensive cyber risk assessment, 2) adaptive nature of cyber risk (see, for example, \citet{DiMase2015}), and 3) correlated losses (see, for example, \citet{Boehme2006, Alderson2014}}

{\color{black}A}nother critical challeng{\color{black}e i}n an enterprise environment is prioritizing and combining different risk management tools, including risk reduction, transfer, and retention, which further leads to {\color{black}budget decisions for implementing the cyber risk management program.} To reduce cyber risk, a company should implement cybersecurity controls, which often involve purchasing cybersecurity solutions and hiring security experts. To transfer the reduced risk, the company should consider buying cyber insurance coverage for specific perils and losses. To retain the reduced but uncovered risk, the company needs emergency funds to mitigate the impact of an incident to ensure business continuity. These risk management options all bear costs. However, the amount of capital that can be budgeted for managing cyber risk is often limited, thus making developing a cost-efficient capital allocation strategy for effective cyber risk management a complex task for enterprises.  

This paper proposes a two-pillar cyber risk management framework that addresses the above challenges in cyber risk assessment and capital allocation. The first pillar, \textit{cyber risk assessment}, utilizes a cascade risk arrival model that unifies the three key components in cybersecurity, including threats, vulnerabilities, and assets, to fully account for the dynamics of cyber risk and the intra-firm loss correlations. The second pillar, \textit{cyber capital management}, balances the trade-off between the marginal costs and marginal benefits of various types of risk management devices and solves for the optimal amounts of capital allocated to them, such that the overall financial implication of cyber risk management is minimized.

Given that cyber risk management is an interdisciplinary subject, the proposed framework takes advantage of approaches to this problem developed by both the cybersecurity community and the actuarial and economics community, which shall be reviewed and compared in the following subsections.


\subsection{Cybersecurity, actuarial, and economics approaches to cyber risk}

Developing innovative risk assessment methods to address the above-mentioned difficulties is interesting to both cybersecurity and risk management communities.  Focusing on different aspects of cyber risks, those two communities have developed vastly different approaches to cyber risk assessment and management.

\subsubsection*{Cybersecurity approach}
In the cybersecurity literature, researchers often focus on the structural properties of cyber systems and study the risks at a microscopic level. For example, 
{\color{black} \citet{Frustaci2018} discussed security issues associated with Internet-of-Things (IoT) systems with extensive insights into the vulnerabilities in individual components of the network, identified that the perception layer of physical sensors is the most vulnerable part of an IoT system, and proposed a score-based risk assessment method for critical security issues}



Based on the same methodology but for more generalized use cases in industrial environments, regulators and leaders of the cybersecurity industry have developed a handful of guidelines and frameworks. For example, the National Institute of Standards and Technology (NIST) has developed a generic set of controls called the  Cybersecurity Framework (CSF) (see \citet{NIST2018}), which can be used by organizations as voluntary guidance for {cyber risk management. S}imilarly, as a cybersecurity industry leader, the Center for Internet Security (CIS) has compiled a detailed list of commonly known vulnerable components in cyber systems and best practices in risk mitigation into CIS Controls (see \citet{CIS2019}). 

In terms of their practicality, one potential shortcoming of these is the lack of insight into the prioritization of tasks from the cost-benefit perspective. For exampl{\color{black}e, NIST CSF sets tiers for cyber risk management practices and encourages organizations to progress from lower-tier practices to higher-tier ones based on cost-benefit analysis. However, how such an analysis can be conducted is missing in CSF.} {\color{black} The lack of economic incentive may discourage organizations from investing in cybersecurity.} As mentioned in \citet{Gordon2015}, private-sector firms have other investment decisions than cybersecurity, and they all compete for limited organizational resources. For this reason, cost-benefit analysis is important in the decision-making process regarding cybersecurity investments. 



Overall, cybersecurity approaches mainly focus on infrastructural details but may not provide enough support for strategic investment and capital allocation planning. The capital component in cyber risk management is largely missing in mainstream solutions provided by the engineering community.

\subsubsection*{Actuarial and economics literature}
In the actuarial literature, more emphasis is on the losses caused by cyber incidents, while cyber systems are often highly abstracte{\color{black}d. 
\citet{Eling2019}} studied a dataset that contains historical cyber losses, identified classes of cyber risks depending on causes, such as human error or technical failure, and then modeled loss frequency and severity by probability distributions. \citet{Eling2018} studied a collection of historical data breaches and used vine copulas to model the dependence among cyber losses in terms of the number of breached records. Using the same dataset of data breaches, \citet{Xu2018} performed trend analyses and built stochastic models for the frequency and size of data breaches caused by hacking. In the economics literature,  \citet{kamiya_risk_2021} combined the information on past data breaches and the impacted firms from multiple sources and showed that the frequency of attacks and the size of cyber losses are influenced by various factors, such as the existence of a risk oversight committee and the type of the breached information. More broadly, some studies focus on the impacts of cyber incidents on the economy. \citet{eisenbach_cyber_2022} showed that the damage caused by cyber-attacks to different banking industry participants could be spread to other entities in the industry through the payments system, thus enlarging the los{\color{black}s. 



O}ther than these studies focusing on the consequences of cyber incidents, there are also several economics studies that try to draw connections between cyber losses and cybersecurity practices.
For example, 
\citet{Gordon2002} expresses the probability of experiencing a security breach as a function of security investment and then uses this probability to calculate the expected loss. The optimal investment in cybersecurity is thus obtained when the marginal investment equals the marginal reduction in expected loss. \citet{kamiya_risk_2021} considers a firm's overall cyber-related cost to be a composite of the expected loss, the cost of compensating stakeholders for cyber risk, and the cost of maintaining a cyber risk management program. Each component depends on the probability of experiencing an attack, and the firm minimizes the overall cost by choosing the optimal attack probability and its corresponding risk management spending. \citet{Smeraldi2014} formulated investment in cybersecurity as {\color{black}a knapsack} optimization proble{\color{black}m
w}here there is a target to protect, a fixed budget, and a set of cyber resources that each can be invested in and generates a return, \textit{i.e.}, some level of protection for the target. The goal is to maximize the overall return with a limited budge{\color{black}t. 
T}hese studies provide some insights into the question of how much should be spent on cybersecurity based on the principle that there should be an equilibrium in the trade-off between cyber investment and the residual cyber ris{\color{black}k. 

T}o generalize this problem {\color{black}further}, we can see it as the trade-off between the immediate cost and the potential cost in the future. This type of problem is {\color{black} widely discussed} in the economics literature on topics other than cyber. For example, the well-known Laffer curve suggests that 
{\color{black}a high tax rate may lead to a high government revenue at present but a low revenue in the future due to reduced economic activities (see \citet{Wanniski1978}.}
Clearl{\color{black}y, 
a balance is needed} between the immediate tax income and the income in the long run {\color{black}so that the overall tax income can be maximized}. Other than taxation, the Laffer curve effect has been studied and applied in many different fields as well; see, for example, \citet{Tragler2001, Janvry2006}. The analog of the Laffer curve is introduced in this paper in the context of cyber risk assessment to explain the trade-off between cybersecurity investment and capitals needed to transfer and absorb the losses.

From a risk perspective, the future benefit could be the reduction in the probability of the occurrence of a loss event. \citet{Menegatti2009} considers a two-period scenario, in which an investment in effort is made today to reduce the probability of a loss tomorrow, and studies how agents with different prudence levels choose their optimal invested efforts. \citet{Hofmann2015} offers an extension to that study, and shows how the curvature of utility and the presence of endogenous saving affect an agent's choice of investing in either reducing the loss probability or reducing the potential loss size.

Ideally, the thought process for investments in cyber should also include the consideration for benefits. However, as found in \citet{Alahmari2020}, organizations rarely incorporate this type of cost-benefit analysis in cyber risk management. What might hinder the development of such strategies is that the existing studies, such as \citet{Gordon2002, Smeraldi2014}, provide no identification of specific cyber assets and vulnerabilities. When multiple vulnerabilities and sources of losses are of concern, the lack of structural information on cyber systems can make it difficult for organizations to create actionable plans and put specific controls in place. 

\subsubsection*{Remarks on comparing classic approaches to frequency, severity, and dependence among cyber losses}
Risks are typically modeled by their frequency and severity in actuarial works, and such attempts have also been made for cyber risk assessment (see, for example, \citet{Xu2018, Eling2019}), where probability distributions are fitted for the occurrences and sizes of cyber losses, and economic insights into the landscape of cyber risks are derived. However, due to the lack of data at a granular level on how cybersecurity investments can impact those distributions, the managerial implications of the classic approach are often limited. Therefore, in our study, the frequency and severity approach is still adopted, as illustrated in Section \ref{sec:case}, but embedded in a comprehensive framework with the structural information of a cyber system that allows for modeling the effect of interventions. 

Similarly, statistical approaches have been used to model the dependence among cyber losses, such as copulas (see, for example, \citet{herath2011copula, peng_modeling_2018, Eling2018}). The utility of these models, especially the parameter-rich ones such as vine copulas, is also largely restricted by the low data availability, and these models can be non-informative at the company level and difficult to guide decision-making with respect to enterprise cyber risk management. Therefore, instead of taking a statistical approach, we utilize the structural information of a cyber system to show how multiple points of failure are logically connected, thus causing correlated losses within a company. This shall be elaborated in Section \ref{sec:risk_assess}. 

\subsection{Capital management framework for cyber risk and contributions}
{\color{black}
To address the shortcomings that 1) some typical cybersecurity approaches to cyber risk management give insufficient consideration to the costs and benefits of cybersecurity, and 2) some typical actuarial approaches lack insights into granular components in cyber systems,   
the proposed framework} quantifies the relationship between the structural information of cyber system components and various kinds of cyber risks that may arise from this system via a cascade risk arrival model. It sets out the amount of risk borne by individual vulnerabilities and assets, and this detailed risk assessment further allows for budgeting decisions regarding cyber risk management. Specifically, this framework optimizes for the amount of capital allocated to various risk management tools, including: 
{\color{black} 1) \textit{ex-ante investments}, which are cybersecurity investments made \textit{before} any incidents happen to reduce the exposure, 2) \textit{cyber insurance}, which partially or entirely transfers risks to insurers, and 3) \textit{ex-post-loss reserves}, which are emergency funds covering residual losses not indemnified by insurance {\it after} the incidents. 
}


These three components represent the reduction, transfer, and retention of cyber risk, respectively, and thus the proposed framework is a comprehensive package for the management of cyber risk. It should be pointed out that the interactions among these risk management approaches have been studied in other scholarly works, such as \citet{eling_optimism_2024}, \citet{bensalem_continuous-time_2023, zeller_risk_2023, bensalem_prevention_2020, wang_integrated_2019,ejrlich_1972}, but the{\color{black}y d}o not consider this problem in a capital allocation settin{\color{black}g. The proposed capital allocation method is made possible by integrating structural information of cyber systems with quantitative risk assessment methods, and this unique feature distinguishes our framework from the aforementioned existing works.} 

Our approach takes all three components into consideration and finds an optimal strategy for businesses to allocate their budgets to implementing various cyber controls and preparing for potential future losses. The goal is to minimize the overall financial implication associated with cyber risk management under a possible budget constraint. To clarify the terminologies in this study, we define \textit{financial implications} as the weighted costs and losses associated with cyber risk management and cyber incidents. In contrast, the direct monetary losses and costs are simply referred to as \textit{costs}. For example, consider the cost of implementing a cybersecurity program to be \$$1$ million. Should a company deem this cost as negligible and assign a small weight of $0.01$ to it, the financial implication of this program would be \$$10,000$. The weights allow risk managers to flexibly prioritize various objectives. This point shall be reiterated as we formulate the capital allocation problem in Section \ref{sec:holistic_allo}. Moreover, the term \textit{benefit} in the notion of cost-benefit analysis refers to the {\it reduction in the financial implication} of cyber risk, and therefore, the cost-benefit analysis we employ involves assessing how changes in direct costs, or simply the costs, can influence financial implications.

{\color{black}The main contribution of this paper is three-fold. First, this paper establishes a unifying framework that integrates the component and structure information of cyber systems with actuarial risk assessment and capital allocation approaches. It is a unique extension of the cascade model in \citet{Bohme2019} and the classic security investment model in \citet{Gordon2002}. Second, the framework is compatible and can be implemented along with the existing standards for cybersecurity, such as the CIS Controls. This feature is important for the practicality of a cyber risk management method but is generally missing in the quantitative methods that have been proposed. Lastly, the paper presents a novel application of collective risk modeling and Pareto optimization, which are rarely discussed in the context of cyber risk. It presents that for cyber risk management in a business environment, there are competing priorities and objectives, and those techniques are well-suited for such a problem.}

The proposed risk assessment and capital allocation methods are implemented in R language, which can be found in the supplementary materials of this paper. Its practicality is demonstrated with real data, applying to one specific company with abundant historical cyber incidents, as well as to a broad spectrum of companies in the same industry. For the former application, practitioners may easily adopt the computer program for their company-specific applications. From the latter application, we are able to draw several general results and insights on cyber risk management; they are that 1) performing cyber scanning is necessary to identify possible attack paths, instead of only relying on historical cyber incidents, 2) the budget constraint of companies with small revenue is generally binding, while companies with large revenue tend to spend a fraction of the budget, and 3) small and medium-sized companies mainly rely on reserves, while large companies generally use diversified strategies with cybersecurity investment, insurance, and reserves together.

The overall organization of the rest of this paper is as follows. Section 2 proposes a tensor-based loss model for cyber risk assessment, which incorporates the structural information of risk arrival. Based on the risk assessment results, Section 3 introduces the holistic framework of optimizing capital allocation for cybersecurity investment, cyber insurance, and loss reserve. Section 4 provides a case study that integrates the risk assessment process and the decision-making of capital allocation for cyber risk management. A discussion of the general insights on cyber risk management is provided in this section as well. Section 5 concludes and outlines some potential future directions.

\section{Cyber Risk Assessment}\label{sec:risk_assess}
\subsection{Cascade model}\label{cascade_model_section}
\citet{Bohme2019} developed a {\it cascade model} to describe the arrival of {a} cyber incident, in which each arrival process is summarized as follows.
\begin{enumerate}
\item[(i)] Once a {\it threat} is initiated by an attacker, it will exploit {\it vulnerabilities} in a cyber system.
\vspace{-2mm}
\item[(ii)] The vulnerabilities that are not fully eliminated by security {\it controls} will then be the viable paths for the threat to impact {\it assets} in the system.
\vspace{-2mm}
\item[(iii)] Eventually, assets of different types and values will generate distinct {\it impacts}.
\end{enumerate}

{This cascade model was developed to explain the process of risk arrival in a cyber system, but was not purposed for the direct quantitative assessment of cyber risks in \citet{Bohme2019}.} In this paper, we tak{e i}nspiration from this model and extend it {to} cyber risk assessment{, and subsequently cyber capital management in the next section}. As summarized in the arrival process, there are five essential cascade components in the model; we shall first explain their details below and us{e e}xample{s} to allude {to} the relationship among them.

A {\it threat} is an action that has the potential {to cause} damage to a cyber system. They might be initiated externally by hackers or cyber criminals with malicious intent, or they could be reckless actions performed internally {that lead} to unintentional damage. For instance, a common threat to password{-}protected accounts is {an} external dictionary attack, which utilizes a large pool of possible passwords to guess the correct one by trial and error until succeeding. {As another example of threats, the spread of viruses may cause correlated losses among multiple organizations (see, for example, \citet{crosignani_pirates_2023}). Given that this paper focuses on managing the cyber risk of a single organization, this scenario of contamination that leads to a global risk correlation can be treated in the same way as other threats from the single organization's perspective.}

A {\it vulnerability} is a cyber weakness that may be exploited by a threat. A threat alone cannot induce any harm if it fails to match any vulnerabilities in the system. For example, allowing an unlimited number of password attempts is a perfect match to the dictionary attack; provided with sufficient computational power, the attacker will eventually get the correct password, which is only a matter of time.

A {\it control} is a security measure that is designed for patching a vulnerability. If a control is applied to a vulnerability, the risk of that vulnerability being exploited by the threat(s) will be reduced. For instance, the compan{\color{black}y l}iable fo{\color{black}r p}assword{-}protected account{\color{black}s c}an limit the number of password attempts before an account is temporarily locked, and hence lengthen th{\color{black}e s}ucceeding time of the attacker; it can further enforce multi-factor authentication, or mandate {a} regular change of passwords to the account holders, as an additional layer of security.

An {\it asset} is associated with the information technology component of a certain entity, which can be tangible or intangible. The value of a cyber asset relies on its confidentiality, integrity, and availabilit{\color{black}y (}see \citet{Cebula2010}), and they are likely to be damaged o{\color{black}r d}estroyed in cyber incidents. For example, a successful password attack would obtai{\color{black}n p}rivate data in a victim{'s} account without {being permitted and n}oticed. Other than physical and digital assets, such as hardware and data, intangible assets, such as reputatio{\color{black}n, c}an also be {\color{black}included} in this cascade model as long as their corresponding impact{\color{black}s c}an be quantified. {\color{black}\citet{gatzert_assessing_2016} extensively discussed how reputation losses might be quantitatively assessed and some available insurance solutions and can be referred to when considering reputation as a type of asset for cyber risk management.} 

An {\it impact} is a materialized loss arising from a cyber asset in a cyber incident. The impact may consist of various types of costs, such as the direct loss of value of a physical cyber asset when it becomes inoperable, or the legal cost if there is a lawsuit derived from the incident. For instance, the successful password attacker might directly thieve any monetary benefit from the victim{'s} account using the data. In addition, companies are obligated to disclose data breach incidents to authorities and customers as required by state laws in the US. Therefore, typical impacts discussed in the literature include direct costs, such as notification costs, regulatory fines, and the indirect cost to shareholder value, which is often used as an indicator of reputation loss (refer to \citet{kamiya_risk_2021}). In this paper, the impact is the base random variable for developing a cyber loss model.

\begin{figure}[ht]
\centering
\includegraphics[width=0.7\textwidth]{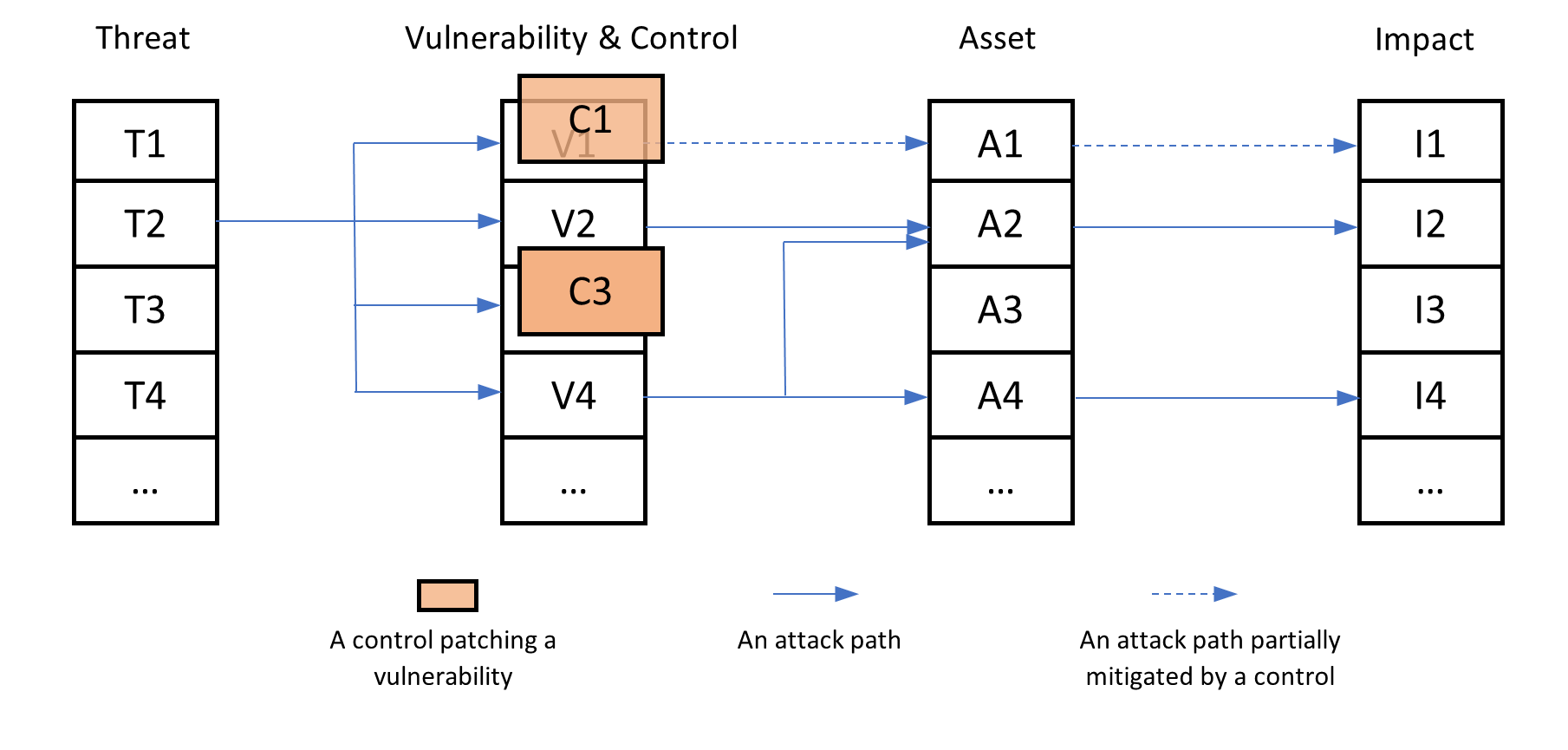}
\caption{Cyber cascade model, and example of cyber incident, due to the second threat, where the first (resp. third) vulnerability is partially (resp. fully) patched by the first (resp. third) control}
\label{fig:los}
\end{figure}

With these five key cascade components being clearly defined, Figure \ref{fig:los} visualizes the structure of the cascade model as well as the relationship among the components. Hereafter, the first four cascade components are occasionally denoted as \textsc{T}, \textsc{V}, \textsc{C}, and \textsc{A}, followed by counting indices; for example, \textsc{T1} represents the first threat while \textsc{V4} represents the fourth vulnerability in the system. Although the impacts are denoted as \textsc{I} in the figure, we shall use another more conventional notation for these base random variables.


\subsection{Mapping relationships among cascade components}\label{sec:map_relation}

To develop a quantitative cyber loss model based on the cascade model proposed by \citet{Bohme2019}, this paper utilizes the relationships among the cascade components to define potential losses. As we can observe from Figure \ref{fig:los}, since an asset being attacked induces one and only one impact, which shall be characterized by a random variable, there are only three mappings to be established, where the vulnerability serves as the core component and relates respectively to the threat, asset, and control. From now on, suppose that there are $l$ types of threats, $m$ categories of vulnerabilities, and $n$ classes of assets.
    \paragraph{\textit{Vulnerability-threat mapping}}
    
    This is a general many-to-many relationship, where a threat may exploit one or more vulnerabilities, and a vulnerability may be exploited by one or more threats. It characterizes the \textit{external risk} by the intrinsic nature of a cyber event, which could be initiated externally or performed internally. This mapping is defined by a matrix $A$, which is of size $l \times m$, such that, for $i = 1,2,\dots,l$ and $j = 1,2,\dots,m$,
    \begin{equation*}
    A_{ij}=
    \begin{cases}
    1 & \text{if \textsc{V}}j \text{ can be exploited by \textsc{T}}i;\\
    0 & \text{otherwise.}
    \end{cases}
    \end{equation*}
    
    \paragraph{\textit{Vulnerability-asset mapping}}
    This is another general many-to-many relationship, where a vulnerability may be associated with one or more assets, and an asset may be associated with one or more vulnerabilities. It characterizes the \textit{internal risk} by the overall configurations inside a cyber system. This mapping is defined by a matrix $B$ of size $m \times n$, such that, for $j = 1,2,\dots,m$ and $k = 1,2,\dots,n$,
    \begin{equation*}
    B_{jk}=
    \begin{cases}
    1 & \text{if \textsc{A}$k$ is associated with \textsc{V}$j$}; \\
    0 & \text{otherwise}.
    \end{cases}
    \end{equation*}
    
    \paragraph{\textit{Vulnerability-control mapping}}
    
    This is naturally a one-to-one and onto relationship based on the definition of controls. It characterizes the effectiveness of each control set in mitigating the corresponding vulnerability. For each vulnerability $j = 1,2,\dots,m$, define $\theta_j\in\left[0,1\right]$ be the scaling factor to the potential loss(es) due to this \textsc{V}$j$, after a certain control set \textsc{C}$j$ is applied.
    When the vulnerability $j$ is only guarded by the control set at the industry standard, $\theta_j=1$. Note that the industry standard, instead of a completely insecure state that no control is applied at all, is considered in this paper, because many basic controls, such as built-in security features of many software applications, exist by default and thus an entirely insecure environment is nowhere to be found. Additional control measure C$j$ reduces the value of $\theta_j$; in particular, when the vulnerability $j$ is fully eliminated, $\theta_j=0$. 
    The vulnerability-control mapping is summarized by the vector $\bm{\theta} = [\theta_1, \theta_2, \dots, \theta_m]$.

Since the vulnerabilities are shared across all three mappings, the relationships among threat, asset, and control can be obtained via composite mappings through the vulnerabilities. These mapping relationships indeed exist in practice, where cybersecurity controls are proposed to patch vulnerabilities and block pathways from threats to assets. For example, in the CIS (Center for Internet Security) Controls framework, \citet{CIS2019} provided a comprehensive relationship between $20$ controls and $5$ assets; for illustration, the relationship is duplicated in Appendix \ref{append:mapping}. Another example is that, based on this framework, \citet{SANSInstitute} did an empirical study, which maps most of those CIS controls with $25$ threats compiled by \citet{VSRCIC2021}; the mapping is also repeated in Appendix \ref{append:mapping2} for demonstration. {\color{black} Guided by CIS Controls or similar standards, companies can inventory their own threats, vulnerabilities, and assets and map out the cascade models corresponding to their security conditions. However, without access to sensitive security information, our study uses a minimal working example introduced in the following subsection to illustrate the proposed framework. In addition, a cascade structure approximated from public records is used for a case study in Section \ref{sec:case}.}

\begin{example}\label{example1}
To plainly illustrate the quantitative cyber loss model, we shall assume the following smaller mapping matrices to be referred repeatedly for risk assessment in the rest of Section \ref{sec:risk_assess}.
\[
A = 
\bordermatrix{%
& \textsc{V1} & \textsc{V2} & \textsc{V3}\cr
\textsc{T1} & 0 & 1 & 0\cr
\textsc{T2} & 0 & 1 & 0\cr
\textsc{T3} & 0 & 1 & 1
}, \quad
B = 
\bordermatrix{%
& \textsc{A1} & \textsc{A2} & \textsc{A3}\cr
\textsc{V1} & 1 & 0 & 1\cr
\textsc{V2} & 1 & 0 & 0\cr
\textsc{V3} & 1 & 1 & 0
}, 
\quad
\bm{\theta} = 
\bordermatrix{%
& \textsc{V1} & \textsc{V2} & \textsc{V3}\cr
& 1/2 & 1/3 & 1/4
}. 
\]
\end{example}

\subsection{Tensor structure}\label{sec:tensor_structure}

With the aid of the three defined mappings above, we can define a single quantity for characterizing the effect by a cyber event on a controlled cyber system. To this end, for $i=1,2,\dots,l$, $j=1,2,\dots,m$, and $k=1,2,\dots,n$, let
\begin{equation}
D_{ijk} = A_{ij} B_{jk} \theta_{j}.
\label{eq:tensor1}
\end{equation}
This quantity summarizes whether the cyber system is in danger of the cyber event, from the $i$-th threat by the $j$-th controlled vulnerability on the $k$-th asset, and if so, how extreme the impact is. To illustrate the intuitions of quantity $D_{ijk}$, we may consider the following three cases.

If $D_{ijk}=1$, the $i$-th threat, $j$-th controlled vulnerability, and $k$-th asset together will induce full impact to the system. This is the aggregate result of that V$j$ is exploited by T$i$ and that A$k$ is associated with V$j$, and when no additional, other than industry standard, control C$j$ is implemented to patch V$j$.

If $0<D_{ijk}<1$, the $i$-th threat, $j$-th controlled vulnerability, and $k$-th asset together will induce partial impact to the cyber system. Note that the partial impact here is defined relative to the full impact when $D_{ijk}=1$, and an impact is considered as partial when additional controls that exceed the industry-average level are implemented but are not yet able to completely eliminate the vulnerability. In this case, V$j$ is exploited by T$i$, A$k$ is associated with V$j$, and C$j$ only partially patches V$j$.

If $D_{ijk}=0$, the $i$-th threat, $j$-th controlled vulnerability, and $k$-th asset together will not induce any impact to the system. This is either because, V$j$ cannot be exploited by T$i$, or A$k$ is not associated with V$j$, or V$j$ is fully patched by C$j$.

The structural information of the cyber cascade model can then be revealed in terms of a {\it tensor} $D$, of size $l\times m\times n$, which shall prove itself useful in developing the quantitative cyber loss model in later sections. To clearly picture this abstract object, we revisit Example \ref{example1} as follows.



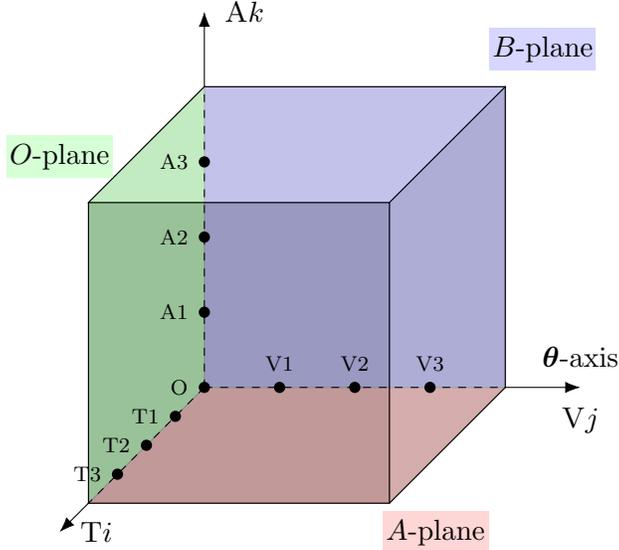
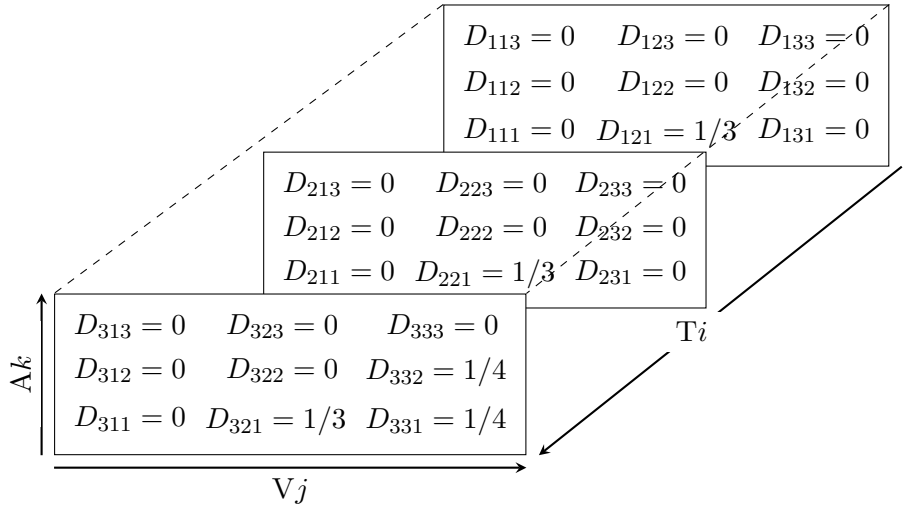
\begin{figure}[!h]
    \centering
    \begin{subfigure}{0.9\textwidth}
    \centering
\begin{tikzpicture}
\coordinate (O) at (0,0,0);
\coordinate (A) at (0,\Width,0);
\coordinate (B) at (0,\Width,\Height);
\coordinate (C) at (0,0,\Height);
\coordinate (D) at (\Depth,0,0);
\coordinate (E) at (\Depth,\Width,0);
\coordinate (F) at (\Depth,\Width,\Height);
\coordinate (G) at (\Depth,0,\Height);
\draw[black,fill=red!80] (O) -- (C) -- (G) -- (D) -- cycle;
\draw[black,fill=blue!80] (O) -- (A) -- (E) -- (D) -- cycle;
\draw[black,fill=green!80] (O) -- (A) -- (B) -- (C) -- cycle;
\draw[black,fill=black!20,opacity=0.8] (D) -- (E) -- (F) -- (G) -- cycle;
\draw[black,fill=black!30,opacity=0.8] (C) -- (B) -- (F) -- (G) -- cycle;
\draw[black,fill=black!10,opacity=0.8] (A) -- (B) -- (F) -- (E) -- cycle;
\draw[dashed] (0,0,0) -- (0,0,4);
\draw[-{Latex[scale=1.2]}] (0,0,4) -- (0,0,5);
\draw[dashed] (0,0,0) -- (0,4,0);
\draw[-{Latex[scale=1.2]}] (0,4,0) -- (0,5,0);
\draw[dashed] (0,0,0) -- (4,0,0);
\draw[-{Latex[scale=1.2]}] (4,0,0) -- (5,0,0);
\node[circle, fill, inner sep = 1.5pt, label=left:{\scriptsize O}] at (0,0,0) {};
\node[circle, fill, inner sep = 1.5pt, label=left:{\scriptsize T1}] at (0,0,1) {};
\node[circle, fill, inner sep = 1.5pt, label=left:{\scriptsize T2}] at (0,0,2) {};
\node[circle, fill, inner sep = 1.5pt, label=left:{\scriptsize T3}] at (0,0,3) {};
\node[circle, fill, inner sep = 1.5pt, label=left:{\scriptsize A1}] at (0,1,0) {};
\node[circle, fill, inner sep = 1.5pt, label=left:{\scriptsize A2}] at (0,2,0) {};
\node[circle, fill, inner sep = 1.5pt, label=left:{\scriptsize A3}] at (0,3,0) {};
\node[circle, fill, inner sep = 1.5pt, label=above:{\scriptsize V1}] at (1,0,0) {};
\node[circle, fill, inner sep = 1.5pt, label=above:{\scriptsize V2}] at (2,0,0) {};
\node[circle, fill, inner sep = 1.5pt, label=above:{\scriptsize V3}] at (3,0,0) {};
\node[label = right:\textsc{T}$i$] at (0,0,5) {};
\node[label = below:\textsc{V}$j$] at (5,0,0) {};
\node[label = right:\textsc{A}$k$] at (0,5,0) {};
\node[label = above:$\bm{\theta}$-axis] at (5,0,0) {};
\node[rectangle, fill, inner xsep = 20pt, inner ysep = 8pt, red!20, opacity = 0.8,
label=center:$A$-plane] at (5,0,5) {};
\node[rectangle, fill, inner xsep = 20pt, inner ysep = 8pt, green!20, opacity = 0.8,
label=center:$O$-plane] at (0,5,5) {};
\node[rectangle, fill, inner xsep = 20pt, inner ysep = 8pt, blue!20, opacity = 0.8,
label=center:$B$-plane] at (4.5,4.5,0) {};
\end{tikzpicture}
    \caption{Tensor structure of cyber cascade model in Example \ref{example1}}
        \vspace{0.5cm}
    \label{fig:cube0}
    \end{subfigure}
    \begin{subfigure}{0.9\textwidth}
    \centering
    \begin{tikzpicture}[every node/.style={anchor=north east,fill=white,minimum width=1.4cm,minimum height=5mm}]
\matrix (mA) [draw, matrix of math nodes, row sep=1pt,
column sep=0.5pt]
{
D_{113} = 0 & D_{123} = 0 \; & D_{133} = 0  \\
D_{112} = 0 & D_{122} = 0 \; & D_{132} = 0  \\
D_{111} = 0 & D_{121} = 1/3 & D_{131} = 0  \\
};

\matrix (mB) [draw,matrix of math nodes] at ($(mA.south west)+(3.5, 0.2)$)
{
D_{213} = 0 & D_{223} = 0 \; & D_{233} = 0  \\
D_{212} = 0 & D_{222} = 0 \; & D_{232} = 0  \\
D_{211} = 0 & D_{221} = 1/3 & D_{231} = 0  \\
};

\matrix (mC) [draw,matrix of math nodes] at ($(mB.south west)+(3.5,0.2)$)
{
D_{313} = 0 & D_{323} = 0 \; & D_{333} = 0 \; \\
D_{312} = 0 & D_{322} = 0 \; & D_{332} = 1/4 \\
D_{311} = 0 & D_{321} = 1/3 & D_{331} = 1/4 \\
};

\draw[thick,-stealth] ([xshift=1ex]mA.south east) -- ([xshift=1ex]mC.south east)
node[midway,xshift=2ex] {\textsc{T}$i$};
\draw[thick,-stealth] ([yshift=-1ex]mC.south west) --
([yshift=-1ex]mC.south east) node[midway,below] {\textsc{V}$j$};
\draw[thick,-stealth] ([xshift=-1ex]mC.south west)
-- ([xshift=-1ex]mC.north west) node[midway,above,rotate=90] {\textsc{A}$k$};
\draw[dashed](mA.north east)--(mC.north east);
\draw[dashed](mA.north west)--(mC.north west);
\end{tikzpicture}
\caption{Elements in tensor of Example \ref{example1}}

\label{fig:cube1}
\end{subfigure}
\caption{Example of tensor representation for cyber cascade model}
\end{figure}

Figure \ref{fig:cube0} depicts the tensor in Example \ref{example1}, while Figure \ref{fig:cube1} summarizes the elements of that tensor. In the example, as there are $3$ types of threats, $3$ categories of vulnerabilities, and $3$ classes of assets, the tensor is defined in the finite set $\{\textsc{T}1, \textsc{T}2, \textsc{T}3\} \times \{\textsc{V}1, \textsc{V}2, \textsc{V}3\} \times \{\textsc{A}1, \textsc{A}2, \textsc{A}3\}$, where $\times$ represents the Cartesian product herein. The tensor can be generalized to include the three mappings by introducing the origin O; indeed, the vulnerability-threat matrix $A$ can be positioned on the $A$-plane, which is defined by the coordinates (T$i$, V$j$, O), while the vulnerability-asset matrix $B$ can be set on the $B$-plane, which is defined by the coordinates (O, V$j$, A$k$); the vulnerability-control vector $\bm{\theta}$ can be put on the $\bm{\theta}$-axis, which is defined by the coordinates (O, V$j$, O). Finally, the $O$-plane, which is defined by the coordinates (T$i$, O, A$k$) and represents the threat-asset mapping matrix $AB$, shall be handy in illustrating the cyber loss model in later sections. Such a generalization also applies for any finite set of threats, vulnerabilities, and assets.

\subsection{Tensor-based cyber loss model} \label{sec:loss_tensor}

As pointed out in Section \ref{cascade_model_section}, an impact from each cyber incident is in fact a base random variable which models a random loss to the controlled cyber system. The impact depends on two factors. The first one is how the cyber event happens; that is, which threat is acted, either internally or externally, and in turn which vulnerability is exploited, as well as which asset is consequently affected; in addition, whether the vulnerability is patched. The first factor has been addressed by the tensor $D$ introduced in the last section, which depends on the controlled cyber system.

The second factor lies in the loss size itself. To this end, for $i=1,2,\dots,l$, $j=1,2,\dots,m$, and $k=1,2,\dots,n$, let $X_{ijk}^{0}$ be the random variable which represents the raw loss due to a cyber incident, from the $i$-th threat by the $j$-th vulnerability on the $k$-th asset. Herein, the raw loss represents a cyber loss which is not further mitigated by additional cybersecurity controls, but is mitigated by those at the industry-average level. 
These random variables can be summarized by another tensor $X^0$, of size $l\times m\times n$.

The contingent impact information from the cyber incident to the controlled cyber system of interest is then given by the element-wise product of the two tensors $D$ and $X^0$; that is,
\begin{equation}
X = D \circ X^0,
\label{eq:tensor2}
\end{equation}
where $\circ$ stands for the element-wise multiplication between two tensors herein. In particular, if the cyber incident is from the $i$-th threat, by the $j$-th controlled vulnerability, on the $k$-th asset, the impact will be given by $X_{ijk}=D_{ijk}X^{0}_{ijk}${; the impacts $X_{ijk}$ would practically be different among various threats, vulnerabilities, and assets.}

{The function of additional cybersecurity controls in this paper is to reduce the impacts from realized incidents, via the vector $\bm{\theta}$. One typical example of such an impact reduction is redundancy, such as database backups, which is not necessarily self-protecting cyber incidents from happening, but assures a speedy and inexpensive recovery. Reducing the probability, of future cyber incidents being emerged, by additional cybersecurity controls, similar to \citet{mazzoccoli_robustness_2020, kamiya_risk_2021, zeller_risk_2023}, shall be incorporated in future studies.}

\begin{example}
We revisit Example \ref{example1} to further illustrate the cyber loss model. While the structure of the tensor in Figure \ref{fig:cube3} of Example \ref{example1} is similar to that in Figure \ref{fig:cube0}, they are actually different since Figure \ref{fig:cube3} depicts the loss tensor in which each element is an impact random variable, as shown in Figure \ref{fig:cube2}. Therein, the elements are calculated based on Equation \eqref{eq:tensor2}. For example, in accordance with Example \ref{example1}, a viable attack path is represented by the red dashed arrows in Figure \ref{fig:cube3}, where T$3$ is acted on V$3$, which is partially patched by C$3$, and consequently affects A$2$. The impact associated with that path is given by
\begin{equation*}
X_{332}=D_{332}X^{0}_{332}=A_{33}B_{32}\theta_{3}X^{0}_{332}=\left(1/4\right)X^{0}_{332}.
\end{equation*}
However, if an attack path is not feasible, such as that T$3$ exploits V$3$ but A$3$ is not associated with V$3$, the impact will become null that
\begin{equation*}
X_{333}=D_{333}X^{0}_{333}=A_{33}B_{33}\theta_{3}X^{0}_{333}=\left(0\right)X^{0}_{333}=0,
\end{equation*}
as $B_{33}=0$, even if the raw random loss $X^{0}_{333}$ is non-zero.
\end{example}

\begin{figure}[!ht]
    \centering
    \begin{subfigure}{0.9\textwidth}
        \centering
\begin{tikzpicture}
    \coordinate (O) at (0,0,0);
    \coordinate (A) at (0,\Width,0);
    \coordinate (B) at (0,\Width,\Height);
    \coordinate (C) at (0,0,\Height);
    \coordinate (D) at (\Depth,0,0);
    \coordinate (E) at (\Depth,\Width,0);
    \coordinate (F) at (\Depth,\Width,\Height);
    \coordinate (G) at (\Depth,0,\Height);
    \draw[black,fill=red!80] (O) -- (C) -- (G) -- (D) -- cycle;
    \draw[black,fill=blue!80] (O) -- (A) -- (E) -- (D) -- cycle;
    \draw[black,fill=green!80] (O) -- (A) -- (B) -- (C) -- cycle;
    \draw[black,fill=black!20,opacity=0.8] (D) -- (E) -- (F) -- (G) -- cycle;
    \draw[black,fill=black!30,opacity=0.8] (C) -- (B) -- (F) -- (G) -- cycle;
    \draw[black,fill=black!10,opacity=0.8] (A) -- (B) -- (F) -- (E) -- cycle;
    \draw[dashed] (0,0,0) -- (0,0,4);
    \draw[-{Latex[scale=1.2]}] (0,0,4) -- (0,0,5);
    \draw[dashed] (0,0,0) -- (0,4,0);
    \draw[-{Latex[scale=1.2]}] (0,4,0) -- (0,5,0);
    \draw[dashed] (0,0,0) -- (4,0,0);
    \draw[-{Latex[scale=1.2]}] (4,0,0) -- (5,0,0);
    \node[circle, fill, inner sep = 1.5pt, label=left:{\scriptsize O}] at (0,0,0) {};
    \node[circle, fill, inner sep = 1.5pt, label=left:{\scriptsize T1}] at (0,0,1) {};
    \node[circle, fill, inner sep = 1.5pt, label=left:{\scriptsize T2}] at (0,0,2) {};
    \node[circle, fill, inner sep = 1.5pt, label=left:{\scriptsize T3}] at (0,0,3) {};
    \node[circle, fill, inner sep = 1.5pt, label=left:{\scriptsize A1}] at (0,1,0) {};
    \node[circle, fill, inner sep = 1.5pt, label=left:{\scriptsize A2}] at (0,2,0) {};
    \node[circle, fill, inner sep = 1.5pt, label=left:{\scriptsize A3}] at (0,3,0) {};
    \node[circle, fill, inner sep = 1.5pt, label=above:{\scriptsize V1}] at (1,0,0) {};
    \node[circle, fill, inner sep = 1.5pt, label=above:{\scriptsize V2}] at (2,0,0) {};
    \node[circle, fill, inner sep = 1.5pt, label=above:{\scriptsize V3}] at (3,0,0) {};
    \node[label = right:\textsc{T}$i$] at (0,0,5) {};
    \node[label = below:\textsc{V}$j$] at (5,0,0) {};
    \node[label = right:\textsc{A}$k$] at (0,5,0) {};
    \node[label = above:$\bm{\theta}$-axis] at (5,0,0) {};
    \node[rectangle, fill, inner xsep = 20pt, inner ysep = 8pt, red!20, opacity = 0.8,
    label=center:$A$-plane] at (5,0,5) {};
    \node[rectangle, fill, inner xsep = 20pt, inner ysep = 8pt, green!20, opacity = 0.8,
    label=center:$O$-plane] at (0,5,5) {};
    \node[rectangle, fill, inner xsep = 20pt, inner ysep = 8pt, blue!20, opacity = 0.8,
    label=center:$B$-plane] at (4.5,4.5,0) {};
    \draw[red, dashed, -{Latex[scale=1.2]}] (0,0,3) -- (3,0,3);
    \draw[red, dashed, -{Latex[scale=1.2]}] (3,0,3) -- (3,2,3);
    \node[circle, fill, inner sep = 1.5pt, label=above left:{\scriptsize (T3, V3)}] at (3,0,3) {};
    \node[circle, fill, inner sep = 1.5pt, label=above left:{\scriptsize(T3, V3, A2)}] at (3,2,3) {};
    \end{tikzpicture}
        \caption{Loss tensor and viable attack path in Example \ref{example1}}
        \vspace{0.5cm}
        \label{fig:cube3}
    \end{subfigure}
    \begin{subfigure}{0.9\textwidth}
        \begin{tikzpicture}[ every node/.style={anchor=north east,fill=white,minimum width=1.4cm,minimum height=5mm}]
\matrix (mA) [draw, matrix of math nodes, row sep=1pt,
column sep=0.5pt]
{
X_{113} = 0 & X_{123} = 0 \qquad & X_{133} = 0  \\
X_{112} = 0 & X_{122} = 0 \qquad & X_{132} = 0  \\
X_{111} = 0 & X_{121} = 1/3 \cdot X_{121}^0 & X_{131} = 0  \\
};

\matrix (mB) [draw,matrix of math nodes] at ($(mA.south west)+(4.5, 0.2)$)
{
X_{213} = 0 & X_{223} = 0 \qquad & X_{233} = 0  \\
X_{212} = 0 & X_{222} = 0 \qquad & X_{232} = 0  \\
X_{211} = 0 & X_{221} = 1/3 \cdot X_{221}^0 & X_{231} = 0  \\
};

\matrix (mC) [draw,matrix of math nodes] at ($(mB.south west)+(4.5,0.2)$)
{
X_{313} = 0 & X_{323} = 0 \qquad & X_{333} = 0 \qquad \\
X_{312} = 0 & X_{322} = 0 \qquad & X_{332} = 1/4 \cdot X_{332}^0  \\
X_{311} = 0 & X_{321} = 1/3 \cdot X_{321}^0 & X_{331} = 1/4 \cdot X_{331}^0  \\
};

\draw[thick,-stealth] ([xshift=1ex]mA.south east) -- ([xshift=1ex]mC.south east)
node[midway, xshift=2ex] {\textsc{T}$i$};
\draw[thick,-stealth] ([yshift=-1ex]mC.south west) --
([yshift=-1ex]mC.south east) node[midway,below] {\textsc{V}$j$};
\draw[thick,-stealth] ([xshift=-1ex]mC.south west)
-- ([xshift=-1ex]mC.north west) node[midway,above,rotate=90] {\textsc{A}$k$};
\draw[dashed](mA.north east)--(mC.north east);
\draw[dashed](mA.north west)--(mC.north west);

\end{tikzpicture}
\caption{Elements in loss tensor of Example \ref{example1}}
\label{fig:cube2}
    \end{subfigure}
\caption{Example of tensor-based cyber loss model}
\end{figure}


\subsection{Aggregate cyber loss} \label{sec:agg_loss}
The abstract tensor-based cyber loss model in the last section does not directly provide much practical use for enterprise risk management purposes. It is, instead, an aggregate cyber loss over a given time period, say one fiscal year, during which multiple cyber incidents might emerge {(for example, a cyber attack might try several times to gain entry into a firm)}, would be called desirable to cyber risk managers. On one hand, as we shall see below, the loss model aggregating across cyber events is not new but based on the well-known collective risk model. On the other hand, the aggregate loss model for each cyber incident is derived by the loss tensor from the cascade model, which does not exist in the literature, and is based on, as alluded in Section \ref{cascade_model_section}, that a cyber event is due to a threat being acted.

\subsubsection{Loss model for each cyber incident}
We first fix a particular cyber event in the given period of time. Let $L$ be the aggregate loss of that cyber incident. Since each incident is initiated by one and only one threat, the aggregate loss is given by the sum of mutually exclusive losses, in which each loss is due to a unique threat. To this end, define, for $i=1,2,\dots,l$,
\begin{equation*}
Y_i=
\begin{cases}
1 & \text{if T}i\text{ is attempted};\\
0 & \text{otherwise},
\end{cases}
\end{equation*}
where, for any distinct $i_1,i_2=1,2,\dots,l$, $\mathbb{P}\left(Y_{i_1}=1,Y_{i_2} = 1\right) = 0$ and $\sum_{i = 1}^l \mathbb{P}\left(Y_i = 1\right) = 1$; denote $\mathbb{P}\left(Y_i = 1\right)$ by $p_i$, and thus $\sum_{i=1}^l p_i = 1$. Moreover, let $Z_i$ be the random loss due to the $i$-th threat, for $i=1,2,\dots,l$; since each threat exploits multiple controlled vulnerabilities, while each vulnerability associates with multiple assets, several impacts would be materialized due to that particular threat being acted; consequently,
\begin{equation*}
Z_i=\sum_{k=1}^{n}\sum_{j=1}^{m}X_{ijk}=\sum_{j=1}^{m}X_{ij1}+\sum_{j=1}^{m}X_{ij2}+\dots+\sum_{j=1}^{m}X_{ijn}.
\end{equation*}
While the order of the double summations does not matter, we choose to aggregate the impacts across vulnerabilities first, and then across assets, to prepare for cyber capital management in later sections, where we shall provide practical reasons for this order. Let also $Z_{ik}$ be the random loss due to the $i$-th threat on the $k$-th asset, for $i=1,2,\dots,l$ and $k=1,2,\dots,n$; and thus,
\begin{equation}
Z_{ik}=\sum_{j=1}^{m}X_{ijk}\quad\text{and}\quad Z_i=\sum_{k=1}^{n}Z_{ik}.
\label{eq:Z_i}
\end{equation}
Therefore, the aggregate loss of the cyber event is given by
\begin{equation}
L=\sum_{i=1}^{l}Z_i\mathds{1}_{\left\{Y_i=1\right\}}.
\label{eq:L}
\end{equation}

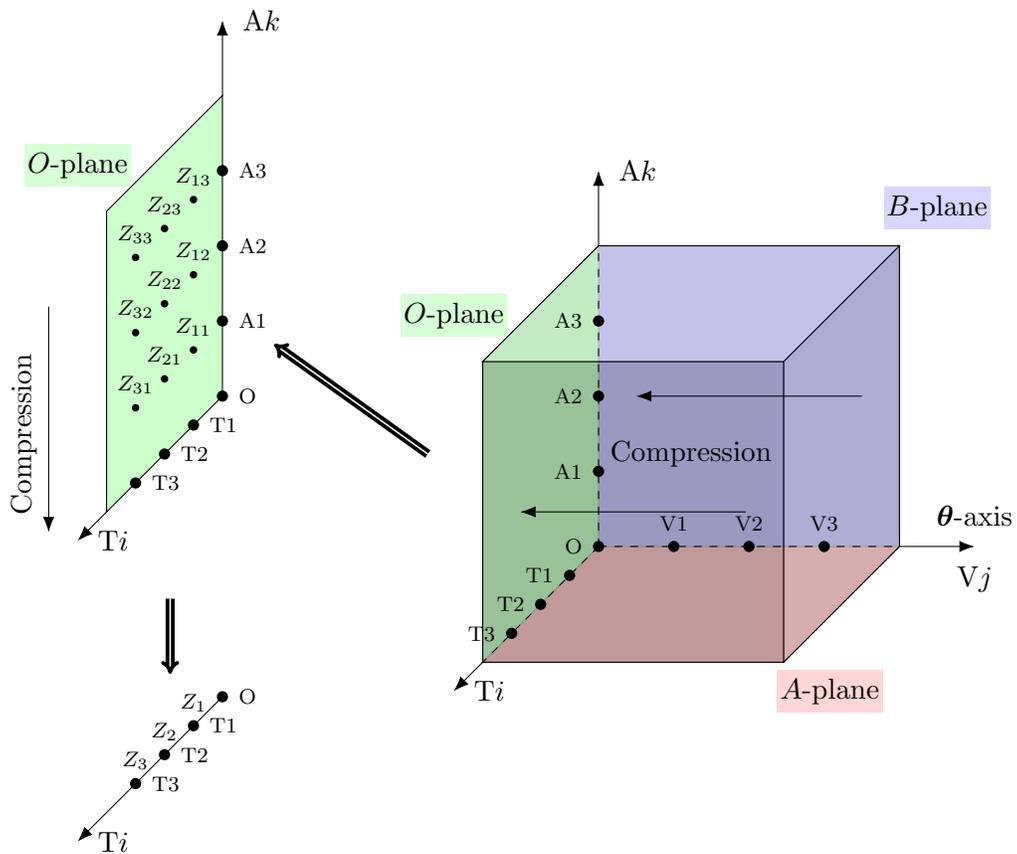
\begin{figure}[ht]
    \centering

\begin{tikzpicture}
\coordinate (O) at (0,0,0);
\coordinate (A) at (0,\Width,0);
\coordinate (B) at (0,\Width,\Height);
\coordinate (C) at (0,0,\Height);
\coordinate (D) at (\Depth,0,0);
\coordinate (E) at (\Depth,\Width,0);
\coordinate (F) at (\Depth,\Width,\Height);
\coordinate (G) at (\Depth,0,\Height);
\draw[black,fill=red!80] (O) -- (C) -- (G) -- (D) -- cycle;
\draw[black,fill=blue!80] (O) -- (A) -- (E) -- (D) -- cycle;
\draw[black,fill=green!80] (O) -- (A) -- (B) -- (C) -- cycle;
\draw[black,fill=black!20,opacity=0.8] (D) -- (E) -- (F) -- (G) -- cycle;
\draw[black,fill=black!30,opacity=0.8] (C) -- (B) -- (F) -- (G) -- cycle;
\draw[black,fill=black!10,opacity=0.8] (A) -- (B) -- (F) -- (E) -- cycle;
\draw[dashed] (0,0,0) -- (0,0,4);
\draw[-{Latex[scale=1.2]}] (0,0,4) -- (0,0,5);
\draw[dashed] (0,0,0) -- (0,4,0);
\draw[-{Latex[scale=1.2]}] (0,4,0) -- (0,5,0);
\draw[dashed] (0,0,0) -- (4,0,0);
\draw[-{Latex[scale=1.2]}] (4,0,0) -- (5,0,0);
\node[circle, fill, inner sep = 1.5pt, label=left:{\scriptsize O}] at (0,0,0) {};
\node[circle, fill, inner sep = 1.5pt, label=left:{\scriptsize T1}] at (0,0,1) {};
\node[circle, fill, inner sep = 1.5pt, label=left:{\scriptsize T2}] at (0,0,2) {};
\node[circle, fill, inner sep = 1.5pt, label=left:{\scriptsize T3}] at (0,0,3) {};
\node[circle, fill, inner sep = 1.5pt, label=left:{\scriptsize A1}] at (0,1,0) {};
\node[circle, fill, inner sep = 1.5pt, label=left:{\scriptsize A2}] at (0,2,0) {};
\node[circle, fill, inner sep = 1.5pt, label=left:{\scriptsize A3}] at (0,3,0) {};
\node[circle, fill, inner sep = 1.5pt, label=above:{\scriptsize V1}] at (1,0,0) {};
\node[circle, fill, inner sep = 1.5pt, label=above:{\scriptsize V2}] at (2,0,0) {};
\node[circle, fill, inner sep = 1.5pt, label=above:{\scriptsize V3}] at (3,0,0) {};
\node[label = right:T$i$] at (0,0,5) {};
\node[label = below:V$j$] at (5,0,0) {};
\node[label = right:A$k$] at (0,5,0) {};
\node[label = above:$\bm{\theta}$-axis] at (5,0,0) {};
\node[rectangle, fill, inner xsep = 20pt, inner ysep = 8pt, red!20, opacity = 0.8,
label=center:$A$-plane] at (5,0,5) {};
\node[rectangle, fill, inner xsep = 20pt, inner ysep = 8pt, green!20, opacity = 0.8,
label=center:$O$-plane] at (0,5,5) {};
\node[rectangle, fill, inner xsep = 20pt, inner ysep = 8pt, blue!20, opacity = 0.8,
label=center:$B$-plane] at (4.5,4.5,0) {};
\draw[-{Latex[scale=1.2]}] (3.5,2,4) -- (0.5,2,4);
\draw[-{Latex[scale=1.2]}] (3.5,2,0) -- (0.5,2,0);
\node[label=center:Compression] at (2,2,2) {}; 
\draw[-{Implies}, double, line width = 0.5mm] (-1.5,2,2) -- (-2,5,6);

\draw[black,fill=green!20]([xshift=-5cm, yshift=2cm]O) -- ([xshift=-5cm, yshift=2cm]A) -- ([xshift=-5cm,yshift=2cm]B) -- ([xshift=-5cm,yshift=2cm]C) -- cycle;
\draw[-{Latex[scale=1.2]}] (-5,2,4) -- (-5,2,5);
\draw[-{Latex[scale=1.2]}] (-5,6,0) -- (-5,7,0);
\node[rectangle, fill, inner xsep = 20pt, inner ysep = 8pt, green!20, opacity = 0.8,
label=center:$O$-plane] at (-5,7,5) {};
\node[circle, fill, inner sep = 1.5pt, label=right:{\scriptsize O}] at (-5,2,0) {};
\foreach \x in {1,...,3}
\foreach \y in {1,...,3} 
 {\pgfmathtruncatemacro{\label}{\y\x}
 \node [circle, fill, inner sep = 1pt, label={above:{\scriptsize $Z_{\label}$}}]  (\x\y) at (-5,\x+2,\y) {};}
\node[label = right:T$i$] at (-5,2,5) {};
\node[label = right:A$k$] at (-5,7,0) {};
\node[circle, fill, inner sep = 1.5pt, label=right:{\scriptsize T1}] at (-5,2,1) {};
\node[circle, fill, inner sep = 1.5pt, label=right:{\scriptsize T2}] at (-5,2,2) {};
\node[circle, fill, inner sep = 1.5pt, label=right:{\scriptsize T3}] at (-5,2,3) {};
\node[circle, fill, inner sep = 1.5pt, label=right:{\scriptsize A1}] at (-5,3,0) {};
\node[circle, fill, inner sep = 1.5pt, label=right:{\scriptsize A2}] at (-5,4,0) {};
\node[circle, fill, inner sep = 1.5pt, label=right:{\scriptsize A3}] at (-5,5,0) {};
\draw[-{Latex[scale=1.2]}] (-5,5.5,6) -- (-5,2.5,6);
\node[label = {[rotate=90, anchor = center]left:Compression}] at (-5,4,6.5) {};
\draw[-{Implies}, double, line width = 0.5mm] (-3,2,7) -- (-3,1,7);

\draw[-{Latex[scale=1.2]}] (-5,-2,0) -- (-5,-2,5);
\node[circle, fill, inner sep = 1.5pt, label=right:{\scriptsize O}] at (-5,-2,0) {};
\node[circle, fill, inner sep = 1.5pt, label=right:{\scriptsize T1}] at (-5,-2,1) {};
\node[circle, fill, inner sep = 1.5pt, label=right:{\scriptsize T2}] at (-5,-2,2) {};
\node[circle, fill, inner sep = 1.5pt, label=right:{\scriptsize T3}] at (-5,-2,3) {};
\foreach \x in {1,...,3}
   {\pgfmathtruncatemacro{\label}{\x}
   \node [circle, fill, inner sep = 1pt, label={above:{\scriptsize $Z_{\label}$}}]  (\x) at (-5,-2,\x) {};}
\node[label = right:T$i$] at (-5,-2,5) {};
\end{tikzpicture}
    
    \caption{Impact aggregation across vulnerabilities and assets in Example \ref{example1}}
    \label{fig:cube4}
\end{figure}

Revisiting Example \ref{example1} should ease the understanding of this series of aggregations. Figure \ref{fig:cube4} illustrates that the impact aggregation across vulnerabilities in \eqref{eq:Z_i} can be pictorially understood as compressing the impact random variables in the loss tensor towards the $O$-plane. Moreover, from Figure \ref{fig:cube4}, the additional impact aggregation across assets in \eqref{eq:Z_i} can be realized as further compressing the aggregated random losses on the $O$-plane towards the T$i$-axis. Finally, the materialized aggregate loss in \eqref{eq:L} due to the cyber incident picks either $Z_1$, $Z_2$, or $Z_3$ on the T$i$-axis.

By \eqref{eq:Z_i} and \eqref{eq:L}, the distribution of aggregate loss of the cyber incident can be obtained by two convolutions, followed by a mixing:
\begin{equation}
f_{Z_{ik}}=f_{X_{i1k}}\ast f_{X_{i2k}}\ast\dots\ast f_{X_{imk}},\quad\text{for $i=1,2,\dots,l$ and $k=1,2,\dots,n$};
\label{eq:Z_ik_dis}
\end{equation}
\begin{equation}
f_{Z_{i}}=f_{Z_{i1}}\ast f_{Z_{i2}}\ast\dots\ast f_{Z_{in}},\quad\text{for $i=1,2,\dots,l$};\quad f_{L}=\sum_{i=1}^{l}p_if_{Z_{i}},
\label{eq:L_dis}
\end{equation}
where $\ast$ represents the convolution operator, and $f_{R}$ is the probability density function of a generic random variable $R$.

\subsubsection{Collective risk model}
Armed with the aggregate loss model for each cyber event, the aggregate loss model over the given period of time {\color{black}simply resembles} the renowned collective risk model. Let $N$ be a non-negative random variable that represents the total number of cyber incidents. Let $L^r$, for $r=1,2,\dots$, be a sequence of independent and identically distributed aggregate losses, where $L^r$ represents the aggregate loss of the $r$-th cyber incident and follows the distribution of $f_L$ given in \eqref{eq:L_dis}. Assume that the total number of cyber incidents $N$ is independent of the aggregate losses $L^r$. Therefore, the {\color{black}collective} loss over the given time period $S=\sum_{r=1}^{N}L^r$, and its probability density function is given by
\begin{equation}
f_S = \lim_{\tilde{n}\rightarrow \infty}
\sum_{r=0}^{\tilde{n}}\left(f_{L^1}\ast f_{L^2}\ast\dots\ast f_{L^r}\right)\times p_{N}\left(r\right),
\label{eq:compound_dis}
\end{equation}
where $p_{N}\left(\cdot\right)$ is the probability mass function of the random variable $N$.

For the purpose of cyber capital management in later sections, let $S_{ik}$ be the {\color{black}collective} loss of cyber incidents from the $i$-th threat on the $k$-th asset over the given time period, where $i=1,2,\dots,l$ and $k=1,2,\dots,n$. It can be expressed as $S_{ik}=\sum_{r=1}^{N_{ik}}Z_{ik}^r$, where $N_{ik}$ is a non-negative random variable to represent the number of cyber incidents from the $i$-th threat on the $k$-th asset, and $Z_{ik}^{r}$, for $r=1,2,\dots$, is a sequence of independent and identically distributed random losses due to T$i$ on A$k$, and each $Z_{ik}^{r}$ represents the random loss of the $r$-th cyber incident associated with (T$i$,A$k$) and follows the distribution of $f_{Z_{ik}}$ given in \eqref{eq:Z_ik_dis}. The distribution of $S_{ik}$ can be obtained in a similar manner as in \eqref{eq:compound_dis}, as long as $N_{ik}$ and $Z_{ik}^r$ are independent. Note that $S=\sum_{i=1}^{l}\sum_{k=1}^{n}S_{ik}$. {\color{black}Again, we} defer providing the practical reasons for this choice of granularity for aggregate loss to the next section.

Recall also that this paper considers additional cybersecurity controls, modeled by the vector $\bm{\theta}$, on only reducing the impacts from the incidents but not on reducing the likelihood of a cyber incident. If the latter case were incorporated, the random variables $N$ and $N_{ik}$, for $i=1,2,\dots,l$ and $k=1,2,\dots,n$, would have been depending on another risk mitigation vector.

\section{Cyber Capital Management} \label{sec:capital_allo}
As in general risk management practice, company managers can make use of the quantitative loss model developed in the {previous} section to outline capital allocation scheme{s} fo{r c}yber risks. {Such schemes should consist of three sequential components: risk reduction, transfer, and retention. 
For risk reduction}, {\it ex-ante investment} should be budgeted {to improve} the existing cybersecurity controls or {implement} new one{s t}o {reduce} potential cyber losses. {For risk transfer, \textit{cyber insurance} coverage should then be purchased by paying a premium to further reduce the exposure. For risk retention}, {\it ex-post-los{s r}eserve} should be planned to weather the {potential residual} cyber losses and to ensure uninterrupted business operations. 

{Existing literature focuses on only one or two of the aforementioned components, but to align well with enterprise risk management principles, a comprehensive and holistic approach to these components should be developed. This section first discusses the roles and interactions of the corresponding capitals for cyber risk reduction, transfer, and retention. It then presents a novel capital allocation model for designing the budgeting decisions on cybersecurity investment, cyber insurance, and reserve based on the proposed cyber risk assessment model in the previous section.}

\subsection{Trade-off: ex-ant{e v}ersus { ex-post-loss capitals}} \label{sec:tradeoff}
Recall that the cyber cascade model is characterized by the three mappings defined in Section \ref{sec:map_relation}. The vulnerability-threat and vulnerability-asset mappings are typically determined by the company's cyber infrastructure and external factors of cyber actions. Company managers have the authority to allocate ex-ante {capital} for cybersecurity {investments} to revise the vulnerability-control mapping. Therefore, an impact, \textit{i.e.}, a loss, from each cyber incident can be {reduced} through partially, or even fully, patching the currently existing vulnerabilities, \textit{i.e.}, any remaining vulnerabilities after industry-standard controls have already been applied, via investment in additional cybersecurity controls.

Let $M_j$, for $j=1,2,\dots,m$, be the cybersecurity investment amount to the control C$j$ for the vulnerability V$j$. The investment amount $M_j$ to the $j$-th control{,} in turn{,} affects the scaling factor to the potential loss(es) due to the V$j$, which is $\theta_j\left(M_j\right)$; in particular, the investment amount and the scaling factor are inversely proportional to each other, that when the ex-ante investment is large, the scaling factor becomes small, and vice versa. Since the total aggregate loss $S$, and the aggregate loss $S_{ik}$ because of (T$i$,A$k$), for $i=1,2,\dots,l$ and $k=1,2,\dots,n$, over the given period of time, depending on the scaling vector $\bm{\theta}$, they are actually driven by the investment vector $\mathbf{M} = \left[M_1,M_2,\dots,M_m\right]$. In the sequel, we shall write such dependence for the aggregate losse{s $S_{ik}\left(\mathbf{M}\right)$}.

\begin{figure}[ht]
   \centering
   \begin{tikzpicture}[scale=1, transform shape, every node/.style={scale=0.7},    bluefill/.style={fill = blue!20},
   redfill/.style={fill=red!20}]

   \draw[->=stealth] (-3.5,0)--(5,0);
   \draw[->=stealth] (-3.5,0)--(-3.5,4) node[above]{\small Total allocated capital} ;
   \node[] at (-3.5,-0.3) {\small Too low};
   \node[] at (4.5,-0.3) {\small Too high};
   \draw (0,1) parabola (4.5,2.8);
   \draw (0,1) parabola (-3.5,3.2);
   \draw[->=stealth, thick] (0.1,2.4) -- (3.5, 2.4);
   \node[align=left] at (1.8, 2.8) {\baselineskip=5pt Excessive investment \\on cybersecurity \par};
   \draw[->=stealth, thick] (-0.1,3.2) -- (-3, 3.2);
   \node[align=right] at (-1.5, 2.8){\baselineskip=5pt Larger impacts \\ due to poor security \par};
   \node[] at (0.5, -0.8) {Allocated {ex-ante} capital fo{r c}ybersecurity investment};
   \draw [blue, fill=blue!20] plot [smooth, tension=1] coordinates { (-3.5,0.1) (0.8,0.7) (4.5,2.7)} -- (4.5, 0) -- (-3.5, 0) -- cycle;
   \draw [red, fill=red!20] plot [smooth, tension=1] coordinates { (-3.5,0.1) (0.8,0.7) (4.5,2.7)} -- (4.5, 2.8) -- (4.5,2.8) parabola bend (0,1) (-3.5,3.2) -- cycle; 
   \draw [dotted] (0, 3.5) -- (0, 0);
   \matrix [draw,below left] at (current bounding box.north east) {
   \node [bluefill,label=right:Ex-ante {capital}] {}; \\
   \node [redfill,label=right:Ex-post{-loss capitals}] {}; \\
};
   \end{tikzpicture}
   \caption{Trade-off between ex-ant{e a}nd ex-post-loss {capitals} for cyber risks}
   \label{fig:cyber_laffer_curve}
\end{figure}
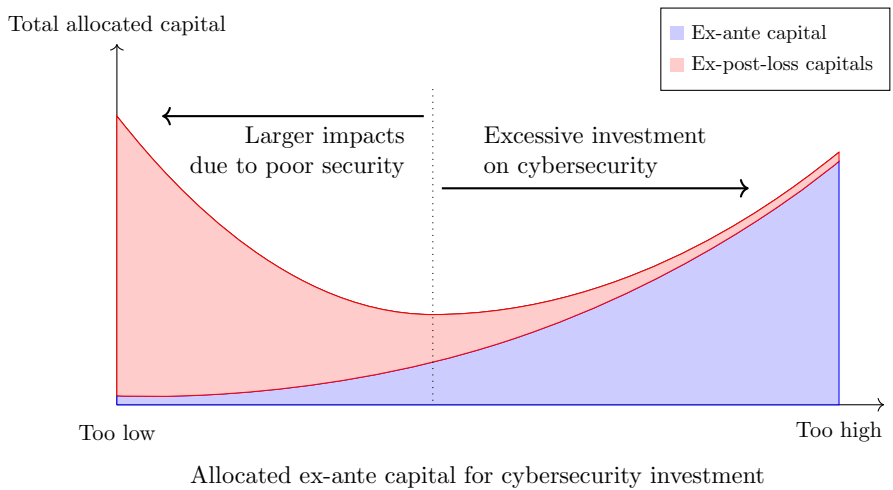

They guide us to the underlying trade-off between ex-ante {capital, which is the cybersecurity investment,} and {ex-post-loss capitals}, {which include the premium for cyber insurance coverage, and the reserve}. When management allocates little or no additional capital for cybersecurity investment beyond the industry standard, the scaling factor vanishes, and the cyber system experiences almost full impact from contingent cyber events. In this case, the realized aggregate losses within the period could be massive, which require{s} the management to plan a tremendous amount of{, premium for sufficient insurance coverage to transfer, and} reserve to absorb losse{\color{black}s. 

H}owever, this does not necessarily imply that the cyber risk manager should fully allocate {ex-ante} capital for cybersecurity investment only. The well-known Gordon-Loeb model (see \citet{Gordon2002, Gordon2016}) states that the marginal benefit, in terms of the {expected loss reduction}, decreases for each unit of cybersecurity investment. Therefore, as pointed out by \citet{Ruan2019}, if the total allocated capital is to be minimized, there will be a certain level of cybersecurity investment such that any additional {ex-ante capital} allocation is not worthy {of} such {a} small marginal return. This corresponds to the right side of the curve in Figure \ref{fig:cyber_laffer_curve}.

Overall, the shape of the curve in Figure \ref{fig:cyber_laffer_curve} resembles the Laffer curve in economics. Both curves reveal {\color{black}the tradeoff between the present and the future costs.}
{\color{black}It also resonates with the optimal risk management investment curve in \citet{kamiya_risk_2021}, which assumes the diminishing marginal benefit of investment, and therefore, the optimal investment is attained when the marginal benefit equals the marginal cost.}

\subsection{Optimal capital allocation}
\label{sec:holistic_allo}


{Finding such} swee{t s}pot for the ex-ant{e a}nd ex-post-loss {capitals} trade-off should be one of the key elements to be considered in any cyber capital management practice. Regarding the {investment decision}, there is no ambiguity that they should be {for} cybersecurity controls that patch vulnerabilities. However, the company could {buy insurance coverage and keep} reserves for different classes of targets. For example, the company could set aside funds for individual assets, \textit{e.g.}, separate reserves for data and physical assets. Similarly, {insurance coverage} could be {purchased} specific{ally for} different threats. In this paper, we follow some existing practices in the cybersecurity an{d c}yber insurance industries. {For instance, the \textit{CyberOne}\textsuperscript{TM} \textit{Coverage} offered by \citet{HSB} covers cyber losses resulting from different attacks, such as malware and denial-of-service attacks, which are various types of threats; see also \citet{chong2023incidentspecific} for incident-specific cyber insurance. The covered losses {in the \textit{CyberOne}\textsuperscript{TM} \textit{Coverage}} include data and system restoration costs, which arise from various classes of assets. {As another example,} cyber risk management tools {and services}, such as the Thrivaca model by \citet{Thrivaca}, offer their clients the estimation of cyber losses at the granularity of threat-asset pairs.} Therefore{, the insurance coverage is written on} the aggregate loss $S_{ik}\left(\mathbf{M}\right)$ due to (T$i$,A$k$), for $i=1,2,\dots,l$, and $k=1,2,\dots,n${, while the reserve is allocated to any retained part of this aggregate loss. This echoes the choice of granularity for aggregate loss in Section \ref{sec:risk_assess}.}

In addition
, there are {three} other common conflicting interests and priorities in a general capital management framework, which we review as follows and shall provide {a} detailed account after formulating the optimal capital allocation problem.


\vspace{-1em}
\paragraph{\textit{Standalone \textit{vs.} corporate allocations}} Naturally, the management has to budget total capital at the whole corporate level which concerns the aggregate los{s.} For project financing, performance measure, and regulatory supervision purposes, the management also has to allocate {capital} at standalone levels. 

\vspace{-1em}
\paragraph{\textit{Cyber risk management goals \textit{vs.} reduction in opportunity cost}} On the one hand, the cyber risk manager should plan and allocate reserves at corporate and standalone levels which do not deviate much from the realized {retained} aggregate losses. On the other hand, reserve needs to be kept as liquid assets, such as cash and money market funds, which typically yield little return; the liquidity requirement causes an opportunity cost, which should be minimized. {\color{black} Similarly, it is in the company's interest to reduce exposure and increase insurance coverage by investing more in cybersecurity and paying more insurance premiums, but these expenses could possibly be used for projects with positive yields, thus also bearing opportunity costs.}

\vspace{-1em}
\paragraph{\textit{Ex-post-loss premium \textit{vs.} reserve allocations}} There is an additional layer of trade-off between the two ex-post-loss capitals. If the cyber risk manager is willing to pay a higher premium to transfer a larger part of the aggregate loss to the insurer, a smaller portion of the loss will be retained, which in turn reduces the needed reserve.

In order to achieve {these three} compromises altogether, we propose to apply the holistic principle by \citet{Chong2021}, which allows the manager to solve the cyber capital allocation as one single optimization problem; see also its recent application to pandemic resources management in \citet{Chen2021pandemic}.

\begin{equation}
\begin{aligned} 
\inf_{\substack{\mathbf{M},{\mathbf{I},}\mathbf{K} \in\mathcal{A}}}&\;  \underbrace{\sum_{j=1}^{m}\eta_j{\color{black}r}M_j}_{\circled{1}}{+\underbrace{\sum_{i=1}^l\sum_{k=1}^n \alpha_{ik}{\color{black}r}\Pi\left[I_{ik}(S_{ik}(\mathbf{M}))\right]}_{\circled{2}}} +\underbrace{\sum_{i=1}^{l}\sum_{k=1}^{n}\nu_{ik}{\color{black}r}K_{ik}}_{\circled{3}} \\ &\; +\underbrace{\sum_{i=1}^{l}\sum_{k=1}^{n}\omega_{ik}\mathbb{E}\left[\left({R_{ik}(S_{ik}(\mathbf{M}))}-K_{ik}\right)^2h_{ik}\left({R_{ik}(S_{ik}(\mathbf{M}))}\right)\right]}_{\circled{4}}\\
&\;+\underbrace{\eta {\color{black}r}M}_{\circled{5}}{+\underbrace{\alpha\sum_{i=1}^l \sum_{k=1}^n{\color{black}r}\Pi\left[ I_{ik}(S_{ik}(\mathbf{M}))\right]}_{\circled{6}}}+\underbrace{\nu {\color{black}r}K}_{\circled{7}}\\ &\; +\underbrace{\omega \mathbb{E}\left[\left({\sum_{i=1}^l \sum_{k=1}^n R_{ik}(S_{ik}(\mathbf{M}))}-K\right)^2h\left({\sum_{i=1}^l \sum_{k=1}^n R_{ik}(S_{ik}(\mathbf{M}))}\right)\right]}_{\circled{8}},\\
\end{aligned}
\label{allo_problem1}
\end{equation}
where
\begin{itemize}
\item $\mathcal{A} = \mathcal{M}\times {\mathscr{F} \times} \mathcal{K} \subseteq \mathbb{R}^m \times {\mathcal{I} \times} \mathbb{R}^{l \times n}$. $\mathcal{M}$ is the set of admissible ex-ante {cybersecurity} investment allocation $\mathbf{M} = \left[M_1,M_2,\dots,M_m\right]$, and $M=\sum_{j=1}^{m}M_j$ is the corporate investment; 
$\mathcal{K}$ is the set of admissible ex-post-loss reserve allocation $\mathbf{K}=\left[K_{ik}\right]_{i=1,2,\dots,l;k=1,2,\dots,n}$, and $K=\sum_{i=1}^{l}\sum_{k=1}^{n}K_{ik}$ is the corporate reserve{; $\mathcal{I}$ is the set of admissible indemnity functions $\mathbf{I} = \left[I_{ik}(\cdot)\right]_{i=1,2,\dots,l;k=1,2,\dots,n}$ for aggregate losses on threat-asset pairs, which shall be defined next, while $\mathscr{F}$ is a subset of $\mathcal{I}$;}
{\item $R_{ik}(\cdot)$ and $I_{ik}(\cdot)$ are respectively the so-called retained and indemnity functions for the aggregate loss $S_{ik}$ on the threat-asset pair (T$i$,A$k$), for $i=1,2,\dots,l$, and $k=1,2,\dots,n$. Note that $R_{ik}(\cdot)+I_{ik}(\cdot) = \mathrm{Id}(\cdot)$, where $\mathrm{Id}(\cdot)$ is the identity function; that is, for a realized loss $s_{ik}(\mathbf{M})$, $R_{ik}(s_{ik}(\mathbf{M}))$ is absorbed by the company, and the other part $I_{ik}(s_{ik}(\mathbf{M}))$ is transferred to the insurer. The admissible set $\mathcal{I}$ of indemnity functions is defined as 
$$\mathcal{I}= \{\left[I_{ik}(\cdot)\right]_{i=1,2,\dots,l;k=1,2,\dots,n} : 0 \le I_{ik}(\cdot) \le \mathrm{Id}(\cdot);\;I_{ik}(\cdot) \text{ and } R_{ik}(\cdot) \text{ are non-decreasing} \},$$
to ensure that the insurance coverage for each aggregate loss $S_{ik}$ satisfies the principle of indemnity and does not lead to any ex-post moral hazards. The functional $\Pi\left(\cdot\right)$ denotes a general premium principle, and thus \circled{2} represents the weighted standalone premiums, in plain dollar unit, paid for coverage to individual threat-asset pairs, and \circled{6} is the total premium, in plain dollar unit, weighted at the corporate level. In the sequel, to simplify the notations, we shall write $I_{ik}(\mathbf{M})$ and $R_{ik}(\mathbf{M})$, for $I_{ik}(S_{ik}(\mathbf{M}))$ and $R_{ik}(S_{ik}(\mathbf{M}))$, as well as $I(\mathbf{M})$ and $R(\mathbf{M})$, for $\sum_{i=1}^l \sum_{k=1}^n I_{ik}(S_{ik}(\mathbf{M}))$ and $\sum_{i=1}^l \sum_{k=1}^n R_{ik}(S_{ik}(\mathbf{M}))$, while $\pi_{ik}(\mathbf{M},I_{ik})$ and $\pi(\mathbf{M}, \mathbf{I})$, for $\Pi\left[I_{ik}(S_{ik}(\mathbf{M}))\right]$ and $\sum_{i=1}^l\sum_{k=1}^n \Pi\left[I_{ik}(S_{ik}(\mathbf{M}))\right]$;
}

\item {\color{black}$r$ is the effective rate of return that this company would be able to earn over the budget period if the capitals were not spent on cybersecurity, cyber insurance, and reserves but on some other project instead at the beginning of the period. It represents the rate of opportunity cost because $r$ multiplied by any of the cybersecurity expenses, premiums, and reserves constitutes an opportunity cost, which can be viewed as a loss to be minimized.}

\item $\eta_j,{\alpha_{ik},} \nu_{ik},\omega_{ik},\eta,{\alpha,}\nu,\omega\geq 0$ are the weights of objectives. Notice that, since the weighted expected deviances, 
$\mathbb{E}\left[\left({R_{ik}(\mathbf{M})}-K_{ik}\right)^2h_{ik}\left({R_{ik}(\mathbf{M})}\right)\right]$
and 
$\mathbb{E}\left[\left({R(\mathbf{M})}-K\right)^2h\left({R(\mathbf{M})}\right)\right]$, are quadratic, their units are measured in square dollars; yet, as the opportunity cost of allocated capitals, $M_j$, $K_{ik}$, $M$, and $K$, are linear, their units are measured in plain dollar{; recall also that (total) premiums are also measured in plain dollar units}. To practically and fairly consider their aggregate effects as in \eqref{allo_problem1}, the weights of the expected deviances, $\omega_{ik}$ and $\omega$, are composed of importance weights with no dollar unit, $\omega_{ik}^{\text{I}}$ and $\omega^{\text{I}}$, and unit-exchange weights with reciprocal dollar units, $\omega_{ik}^{\text{E}}$ and $\omega^{\text{E}}$, that each weight is their products, \textit{i.e.}, $\omega_{ik} = \omega_{ik}^\text{I}\omega_{ik}^\text{E}$ for $i = 1,2,\dots,l$ and $k = 1,2,\dots,n$, and $\omega = \omega^\text{I}\omega^\text{E}$. The importance weights, $\omega_{ik}^{\text{I}}$ and $\omega^{\text{I}}$, are compared relatively with other importance weights of no dollar unit, $\eta_j$, {$\alpha_{ik}$,} $\nu_{ik}$, $\eta$, {$\alpha$,} and $\nu$. The choice of unit-exchange weights, $\omega_{ik}^{\text{E}}$ and $\omega^{\text{E}}$, shall be discussed with an aid of the numerical example in Section \ref{sec:case};
\item $h_{ik}\left({\cdot}\right)$ and $h\left({\cdot}\right)$ are the weighted penalty functions, for which $\mathbb{E}\left[h_{ik}\left(S_{ik}\right)\right]=\mathbb{E}\left[h\left(S\right)\right]=1$, which measure the deviation of realized {retained} aggregate losses from the allocated reserve intended to cover the losses; the family of such functions is summarized in \citet{Furman2008, Dhaene2012, Chong2021}.
\end{itemize}

The holistic cyber capital allocation solved by \eqref{allo_problem1} shall indeed take those three equilibriums into account, which can also be read from various pieces in the objective function.

\vspace{-1em}
\paragraph{\textit{Standalone and corporate allocations}} The standalone allocation scheme is represented by $\circled{1}${, $\circled{2}$, $\circled{3}$, and $\circled{4}$}, where investments on cybersecurity controls{, transferred losses,} and reserve{s a}re balanced on the level of vulnerabilities and threat-asset pairs. In contrast, {$\circled{5}$, $\circled{6}$, $\circled{7}$, and $\circled{8}$} together represent the corporate allocation scheme, in which the total investment{, insured loss,} and reserve are determined in the equilibrium.

\vspace{-1em}
\paragraph{\textit{Loss-reserve matching and reduction in opportunity cost}} The matchings are given in $\circled{4}$ for standalone allocations, and in $\circled{8}$ for corporate allocation. They are achieved by minimizing the mismatching cost, which is the weighted expected deviance between the realized {retained} aggregate loss and the allocated reserve for covering the loss. The deviance is given in the quadratic form since cyber incidents might cause ripple effects, as described in \citet{UniOxfordAxis2020}, which could lead to additional losses if the initial impact cannot be well contained due to the lack of reserve. Therefore, the relationship between loss-reserve mismatching and its cost should be non-linear. For instance, if a cyber attack disrupts the production lines of a manufacturer, and there is not sufficient fund {nor insurance coverage} for resuming the operation promptly, the manufacturer may face additional contractual liabilities for the late delivery of products. As for minimizing the opportunity cost for holding liquid assets as reserve, they are represented by $\circled{3}$ for standalone allocations, and by $\circled{7}$ for corporate allocation.

\vspace{-1em}
\paragraph{\textit{Ex-post-loss premium and reserve allocations}} The cost of standalone premiums is given by their weighted sum $\circled{2}$, while the counterpart in the corporate level is given by $\circled{6}$. The trade-off between the two ex-post-loss capitals, which are the premium and reserve, discussed in Section \ref{sec:holistic_allo} can be revealed in \eqref{allo_problem1}. Should the insurance coverage be larger, the retained aggregate losses of the company are reduced, and thus, by $\circled{4}$ and $\circled{8}$, the needed reserves matching for the losses, as well as their {\color{black}weighted} opportunity costs given in $\circled{3}$ and $\circled{7}$, are also lowered; yet, the insurer should charge a higher premium, and hence the {\color{black}weighted} cost of premiums are increased in $\circled{2}$ and $\circled{6}$.

\vspace{-1em}
\paragraph{\textit{Ex-ante and ex-post-loss capitals}} The trade-off between {ex-ante} and {ex-post-loss capitals} introduced in Section \ref{sec:tradeoff} is also weighed in. The {\color{black}weighted} costs of cybersecurity investment are given in $\circled{1}$ for standalone allocations, and in $\circled{5}$ for corporate allocation, whereas the {\color{black}weighted} standalone and corporate insurance coverage and reserve costs are $\circled{2}$, $\circled{3}$, and $\circled{4}$ together, as well as $\circled{6}$, $\circled{7}$, and $\circled{8}$ together, respectively. Note that the cost of reserve incorporates the consideration for the balancing between sufficient loss reserve and economical opportunity cost, as discussed in the second point above.

{\color{black}The eight pieces in \eqref{allo_problem1} are weighted costs representing the relative importance of various objectives, while the direct and unweighted costs, including the upfront cybersecurity expenses, reserves, and insurance premiums, can be impacted by the budget for cyber risk management. For clarity, we differentiate between weighted and unweighted costs and refer to the unweighted costs as {\it costs} and the weighted costs as {\it financial implications} hereafter.}

\subsection{Holistic cyber capital allocation}\label{sec:holistic}

The capital allocation problem in \eqref{allo_problem1} can be solved in two steps. First, for each {cybersecurity} investment $\mathbf{M}\in\mathcal{M}$ {and for each combination of indemnity functions $\mathbf{I} \in \mathscr{F}$}, the optimal ex-post-loss reserve allocation $\mathbf{K}^*\in\mathcal{K}$, which depends on the fixed investment $\mathbf{M}$ {and indemnity functions $\mathbf{I}$}, is solved from \eqref{allo_problem1} without the objective terms {$\circled{1}$, $\circled{2}$, $\circled{5}$, and $\circled{6}$}. Then, the optimal reserve {$\mathbf{K}^*\left(\mathbf{M}, \mathbf{I}\right)$} is substituted back into \eqref{allo_problem1}, which is solved for the optimal investment $\mathbf{M}^*$ {and optimal indemnities $\mathbf{I}^*$}. Consequently, the optimal {cybersecurity} investment{, insurance coverage,} and reserve allocation {are} given by the {tuple $\left(\mathbf{M}^*,\mathbf{I}^*, \mathbf{K}^*\left(\mathbf{M}^*, \mathbf{I}^*\right)\right)$}. The first step was partially resolved in 
\citet{Chong2021}, while we shall herein impose an additional total budget constraint for {cybersecurity} investments{, premiums,} and reserves together. As pointed out in \citet{Smeraldi2014}, most organizations do not have the funds to implement strong security, and thus cyber risk management decisions can only be made with a limited budget in most cases.
For the optimal investment $\mathbf{M}^*$ {and insurance $\mathbf{I}^*$} in the second step, a pairwise comparison between admissible investment allocations{, and between admissible insurance coverages,} shall again shed light on the trade-off{s, between ex-post-loss premium and reserve allocations, and between ex-ante and ex-post-loss capitals}. In the sequel, an investment or a reserve is admissible as long as it is non-negative, which should be practically assumed. 

\subsubsection{Optimal ex-post-loss reserve}
Throughout this subsection, fix a non-negative investment allocation $\mathbf{M}$ {and admissible indemnities $\mathbf{I}$,} such that $M {+ \pi(\mathbf{M}, \mathbf{I})} \leq\beta$, where $\beta$ is the total budget for {cybersecurity} investment{, premium for insurance,} and reserve allocations{. 

W}e shall first present the unconstrained solution, {\it i.e.}, when $\mathcal{K}=\mathbb{R}^{l\times n}$ in \eqref{allo_problem1}, as it shall lead us on how to solve the optimal reserve with { constraints. In this paper, two sets of constraints shall be considered for practicality. We first show how reserves should be determined if they are only subject to the non-negativity requirement, and then, in addition, a total budget constraint is assumed.}

\paragraph{Unconstrained case}
With $\mathcal{K}=\mathbb{R}^{l\times n}$, the optimal ex-post-loss reserve {$\widetilde{\mathbf{K}}^*\left(\mathbf{M}, \mathbf{I}\right)$} is given by, for any $i'=1,2,\dots,l$, and $k'=1,2,\dots,n$,
\begin{equation}
{\widetilde{K}^*_{i'k'}\left(\mathbf{M},\mathbf{I}\right)}= {\overline{K}_{i'k'}\left(\mathbf{M},\mathbf{I}\right)}- \widetilde{W}_{i'k'}\left(\sum_{i=1}^l\sum_{k=1}^n {\overline{K}_{ik}\left(\mathbf{M},\mathbf{I}\right)} - {\overline{K}\left(\mathbf{M},\mathbf{I}\right)}\right),
\label{eq:holistic_unconstrained}
\end{equation}
and the corporate reserve is
\begin{equation*}
{\widetilde{K}^*\left(\mathbf{M},\mathbf{I}\right)}=\sum_{i=1}^{l}\sum_{k=1}^{n}{\widetilde{K}^*_{ik}\left(\mathbf{M},\mathbf{I}\right)}={\overline{K}\left(\mathbf{M},\mathbf{I}\right)}+\widetilde{W}\left(\sum_{i=1}^l \sum_{k = 1}^n {\overline{K}_{ik}\left(\mathbf{M},\mathbf{I}\right)} - {\overline{K}\left(\mathbf{M},\mathbf{I}\right)}\right),
\end{equation*}
where, for any $i'=1,2,\dots,l$, and $k'=1,2,\dots,n$,
\begin{equation}
{\overline{K}_{i'k'}\left(\mathbf{M},\mathbf{I}\right)} = \mathbb{E}\left[{R_{i'k'}(\mathbf{M})}h_{i'k'}\left({R_{i'k'}(\mathbf{M})}\right)\right] - \frac{{\color{black}r}\nu_{i'k'}}{2\omega_{i'k'}},\quad {\overline{K}\left(\mathbf{M},\mathbf{I}\right)} = \mathbb{E}\left[{R(\mathbf{M})}h\left({R(\mathbf{M})}\right)\right] - \frac{{\color{black}r}\nu}{2\omega},
\label{eq:individual_components}
\end{equation}
\begin{equation*}
\widetilde{W}_{i'k'} = \frac{\frac{1}{\omega_{i'k'}}}{\frac{1}{\omega}+\sum_{i=1}^l \sum_{k=1}^n \frac{1}{\omega_{ik}}},\quad \widetilde{W} = \frac{\frac{1}{\omega}}{\frac{1}{\omega} + \sum_{i=1}^l \sum_{k=1}^n \frac{1}{\omega_{ik}}};
\end{equation*}
for its proof, see \citet{Chong2021}. These four components constituting the optimal and corporate reserves are respectively the optimal standalone reserve {$\overline{K}_{i'k'}\left(\mathbf{M}, \mathbf{I}\right)$}, the optimal corporate reserve {$\overline{K}\left(\mathbf{M},\mathbf{I}\right)$}, and harmonic weights $\widetilde{W}_{i'k'}$, $\widetilde{W}$. Note that $\widetilde{W}+\sum_{i=1}^{l}\sum_{k=1}^{n}\widetilde{W}_{ik}=1$. The optimal standalone reserve describes the amount planned for a particular threat-asset pair $\left(\text{T}i',\text{A}k'\right)$ if it were the only pair to be allocated with reserve. When ${\color{black}r}\nu_{i'k'}$ is relatively larger than $\omega_{i'k'}$, {\it i.e.}, it is more costly to allocate reserve for the threat-asset pair $\left(\text{T}i',\text{A}k'\right)$, in terms of the opportunity cost over the loss-reserve mismatching cost, its optimal reserve is reduced. Similar interpretations hold for the optimal corporate reserve. The harmonic weights represent the competition among optimal standalone reserves and the optimal corporate reserve. When $\omega_{i'k'}$ is relatively larger than the other $\omega_{ik}$ and $\omega$, the corresponding harmonic weight $\widetilde{W}_{i'k'}$ is small, and thus the optimal reserve inclines to its own optimal standalone reserve.
\vspace{-1em}
\paragraph{Constrained case: non-negative reserves} Let us discuss the case when the non-negativity constraint, {\it i.e.}, $\mathcal{K}=\mathbb{R}^{l\times n}_+$, is imposed. On {the} one hand, if all of the unconstrained optimal ex-post-loss reserve{s} solved in \eqref{eq:holistic_unconstrained} satisfy the constraint, they are also optimal for the constrained case. On the other hand, if some of the unconstrained reserve{s} solved in \eqref{eq:holistic_unconstrained} do not satisfy the constraint, they should instead be assigned by the lower bound {of a} reserve, which is zero in this setting. Hence, these threat-asset pairs should not compete with other pairs, as well as the whole corporate, for reserve resources, and this should be reflected in the revised harmonic weights{,} which exclude those non-competing threat-asset pairs. However, before all of the constrained optimal ex-post-loss reserve{s} are solved, the exact threat-asset pairs being bound by the non-negativity constraint are {\it not known a priori}. To this end, we introduce an auxiliary variable {$\delta_{ik}$}, for $i=1,2,\dots,l$, and $k=1,2,\dots,n$, which takes a value either $0$ or $1$. Then, all of the constrained optimal ex-post-loss reserves {$K^+_{i'k'}\left(\mathbf{M}, \mathbf{I}\right)$}, together with the auxiliary variables as by-products, are solved by the following set of equations.
\begin{equation}
{K^+_{i'k'}\left(\mathbf{M}, \mathbf{I}\right)}=
\begin{cases}
{K_{i'k'}\left(\mathbf{M}, \mathbf{I}\right)} &\text{if}\quad {K_{i'k'}\left(\mathbf{M}, \mathbf{I}\right)}\geq 0\\
0 &\text{if}\quad {K_{i'k'}\left(\mathbf{M}, \mathbf{I}\right)}< 0
\end{cases},\quad \mathds{1}_{\left\{{K_{i'k'}\left(\mathbf{M}, \mathbf{I}\right)}\geq 0\right\}}={\delta_{i'k'}},
\label{eq:holistic_constrained}
\end{equation}
for all $i'=1,2,\dots,l$, and $k'=1,2,\dots,n$, where
\begin{equation*}
{K_{i'k'}\left(\mathbf{M}, \mathbf{I}\right)}= {\overline{K}_{i'k'}\left(\mathbf{M}, \mathbf{I}\right)}- W_{i'k'}^+\left(\sum_{i=1,i\neq i'}^l\sum_{k=1,k\neq k'}^n {\overline{K}_{ik}\left(\mathbf{M}, \mathbf{I}\right)\delta_{ik}}+{\overline{K}_{i'k'}\left(\mathbf{M}, \mathbf{I}\right)}- {\overline{K}\left(\mathbf{M}, \mathbf{I}\right)}\right),
\end{equation*}
and the optimal standalone reserve {$\overline{K}_{i'k'}\left(\mathbf{M}, \mathbf{I}\right)$}, as well as the optimal corporate reserve {$\overline{K}\left(\mathbf{M}, \mathbf{I}\right)$}, are still given by \eqref{eq:individual_components}, but the harmonic weight is revised as
\begin{equation*}
W_{i'k'}^+=\frac{\frac{1}{\omega_{i'k'}}}{\frac{1}{\omega}+\frac{1}{\omega_{i'k'}}+\sum_{i=1,i\neq i'}^l \sum_{k=1,k\neq k'}^n \frac{1}{\omega_{ik}}{\delta_{ik}}}.
\end{equation*}
For a further detailed account of the constrained solution and the proof of obtaining the set of equations, see \citet{Chong2021}.
\vspace{-1em}
\paragraph{Constrained case: non-negative reserves with total budget}{ For many businesses, there could only be a limited amount of capital devoted to cyber risk management, which may not be sufficient to meet the reserving scheme given by Equation \eqref{eq:holistic_constrained}. {\color{black}Acknowledging that the primary goal of businesses is to generate profit, the proposed framework treats the budget for cyber risk management as an exogenous parameter that can be flexibly chosen by the company so as not to compromise its general goal of profit-seeking.} Since $\beta\geq 0$ is the total budget for cybersecurity investment, premium for insurance, and reserve allocations, for the fixed investment allocations $\mathbf{M}$ {and insurance coverages $\mathbf{I}$}, the remaining budget for ex-post-loss reserves is ${b\left(\mathbf{M}, \mathbf{I}\right)}:=\beta-M{-\pi(\mathbf{M}, \mathbf{I})}$. Together with the non-negative reserve constraint, $\mathcal{K} = \mathbb{R}^{l\times n}_+ \cap \left\{{\mathbf{K}(\mathbf{M}, \mathbf{I})} \; \middle| \; \sum_{i = 1}^l \sum_{k = 1}^n {K_{ik}(\mathbf{M}, \mathbf{I})} \le {b\left(\mathbf{M}, \mathbf{I}\right)}\right\}$.


Let {$K_{i'k'}^{*}\left(\mathbf{M}, \mathbf{I}\right)$}, for $i' = 1,2,\dots, l$ and $k'=1,2,\dots, n$, be the solution in this scenario. The total budget constraint is not binding if $\sum_{i=1}^l\sum_{k=1}^n {K^+_{ik}(\mathbf{M}, \mathbf{I})}\leq {b\left(\mathbf{M}, \mathbf{I}\right)}$, which essentially results in the solution given by Equation \eqref{eq:holistic_constrained}, \textit{i.e.}, {$K_{i'k'}^{*}\left(\mathbf{M}, \mathbf{I}\right) = K_{i'k'}^{+}\left(\mathbf{M}, \mathbf{I}\right)$}. Therefore, we shall only consider the case in which the total budget for reserves is smaller than $\sum_{i=1}^l\sum_{k=1}^n {K^+_{ik}(\mathbf{M}, \mathbf{I})}$.

By the Karush–Kuhn–Tucker conditions, the optimal ex-post-loss reserves under the binding budget constraint, {$K_{i'k'}^{*}\left(\mathbf{M}, \mathbf{I}\right)$}, can be expressed as follows,
\begin{equation}
{K^{*}_{i'k'}\left(\mathbf{M}, \mathbf{I}\right)}=
\begin{cases}
{\overline{K}_{i'k'}\left(\mathbf{M}, \mathbf{I}\right)} - W_{i'k'}\left(\sum_{(i,k) \in {\mathfrak{I}}} {\overline{K}_{ik}\left(\mathbf{M}, \mathbf{I}\right)} - {b\left(\mathbf{M}, \mathbf{I}\right)} \right) &\text{if}\quad (i', k') \in {\mathfrak{I}}\\
0 &\text{if}\quad (i', k') \notin {\mathfrak{I}}
\end{cases},
\label{eq:holistic_budget_constrained}
\end{equation}
where $W_{i'k'}$ is the harmonic weight given by $W_{i'k'} = \frac{1}{\omega_{i'k'}}/\sum_{(i, k) \in {\mathfrak{I}}} \frac{1}{\omega_{ik}}$, and the set of indices {$\mathfrak{I}$} is a subset of $\left\{1,2,\dots,l\right\}\times \left\{1,2,\dots,n\right\}$ which satisfies:
$$
{\mathfrak{I}} = \left\{\left(i', k'\right) \;  \middle| \; i' = 1,\dots, l; k' = 1,\dots, n{; b\left(\mathbf{M}, \mathbf{I}\right)}+\sum_{(i, k) \in {\mathfrak{I}}} \left(\frac{\omega_{i'k'}}{\omega_{ik}}{\overline{K}_{i'k'}\left(\mathbf{M}, \mathbf{I}\right)}   - {\overline{K}_{ik}\left(\mathbf{M}, \mathbf{I}\right)}\right) \ge 0 \right\}.
$$
Appendix \ref{append:algo} provides {the} pseudo-code to identify the set {$\mathfrak{I}$}.
}

\subsubsection{Optimal {cybersecurity} investment {and insurance coverage}}\label{sec:investment_section}
With the {optimized} reserv{e g}iven by \eqref{eq:holistic_budget_constrained}, the total cost \eqref{allo_problem1}, which depends on the ex-ante investment for cybersecurity controls {and insurance coverage for aggreggate losses on threat-asset pairs}, involves {three} parts:
\begin{itemize}
\item the residual {\color{black}financial implication} of reserve
\begin{equation}
\begin{aligned}
{g_r\left(\mathbf{M}, \mathbf{I}\right)}=&\;\sum_{i=1}^{l}\sum_{k=1}^{n}{\color{black}r}\nu_{ik}{K^*_{ik}\left(\mathbf{M}, \mathbf{I}\right)}\\&\;+\sum_{i=1}^{l}\sum_{k=1}^{n}\omega_{ik}\mathbb{E}\left[\left({R_{ik}(\mathbf{M})}-{K^*_{ik}\left(\mathbf{M}, \mathbf{I}\right)}\right)^2h_{ik}\left({R_{ik}(\mathbf{M})}\right)\right]\\&\;+{\color{black}r}\nu {K^*\left(\mathbf{M}, \mathbf{I}\right)}+\omega \mathbb{E}\left[\left({R(\mathbf{M})}-{K^*\left(\mathbf{M}, \mathbf{I}\right)}\right)^2h\left({R(\mathbf{M})}\right)\right];
\end{aligned}
\label{eq:g_r}
\end{equation}
\item the {\color{black}financial implication} of cybersecurity investment
\begin{equation}
g_c\left(\mathbf{M}\right)=\sum_{j=1}^{m}\eta_j{\color{black}r}M_j+\eta {\color{black}r}M;
\label{eq:g_c}
\end{equation}
{\item the {\color{black}financial implication} of insurance premiums
\begin{equation}
    g_{I}(\mathbf{M}, \mathbf{I}) = \sum_{i=1}^l\sum_{k=1}^n \alpha_{ik}{\color{black}r}\pi_{ik}(\mathbf{M},I_{ik}) + \alpha{\color{black}r}\pi(\mathbf{M}, \mathbf{I}).
    \label{eq:g_i}
\end{equation}}
\end{itemize}
{Via these, the trade-offs, between ex-post-loss premium and reserve allocations, and between ex-ante and ex-post-loss capitals, could be crisply represented. Generally speaking, both are due to the fact that the marginal benefit outweighs the marginal {\color{black}financial implication}, which is well-known in economics as the cost-benefit analysis.

For the former one, fix any admissible cybersecurity investment $\mathbf{M}\in\mathcal
{M}$. For any two admissible $\mathbf{I}_p,\mathbf{I}_q\in\mathscr{F}$ such that $M + \pi(\mathbf{M}, \mathbf{I}_p)\leq \beta$ and $M + \pi(\mathbf{M}, \mathbf{I}_q)\leq \beta$, the insurance coverage $\mathbf{I}_p$ is better off than the other insurance coverage $\mathbf{I}_q$ if
\begin{equation*}
g_r\left(\mathbf{M}, \mathbf{I}_q\right)-g_r\left(\mathbf{M}, \mathbf{I}_p\right)\geq g_{I}(\mathbf{M}, \mathbf{I}_p)-g_{I}(\mathbf{M}, \mathbf{I}_q).
\end{equation*}
The inequality entails that, if the reduction of the residual {\color{black}financial implication} of reserve is more than the additional {\color{black}financial implication} of insurance premiums, the exploring insurance coverage $\mathbf{I}_p$ is worth more than the existing insurance coverage $\mathbf{I}_q$.

For the latter one, similarly, fix any admissible insurance coverage $\mathbf{I}\in\mathscr{F}$. For any two admissible $\mathbf{M}_p,\mathbf{M}_q\in\mathcal{M}$ such that $M_p + \pi(\mathbf{M}_p, \mathbf{I})\leq \beta$ and $M_q + \pi(\mathbf{M}_q, \mathbf{I})\leq \beta$, the exploring cybersecurity investment $\mathbf{M}_p$ is better off than the existing cybersecurity investment $\mathbf{M}_q$ if
\begin{equation*}
g_r\left(\mathbf{M}_q, \mathbf{I}\right)-g_r\left(\mathbf{M}_p, \mathbf{I}\right)\geq g_c\left(\mathbf{M}_p\right)-g_c\left(\mathbf{M}_q\right),
\end{equation*}
as the reduction of the residual {\color{black}financial implication} of reserve is more than the additional {\color{black}financial implication} of cybersecurity investment.} This is perfectly in line with the advocate in cybersecurity investment literature, such as \citet{Cavusoglu2004,Gordon2002, Ruan2019}, that an investment should not be made if it does not generate more benefit than the investment itself.

Two considerations for practicality are worth mentioning due to the possible restrictions on a company's abilities to choose cybersecurity investments and insurance coverage.

Firstly, a company might be obligated to adhere to specific cybersecurity regulations, limiting its flexibility in opting out of certain security controls. For example, a utility company needs to comply with NERC CIP standards (see \citet{NERC}) and periodically provide security awareness training to its personnel. In that case, all feasible $\mathbf{M}$ should include the required investments, and those cybersecurity investments lower the corresponding elements in the scaling vector $\bm{\theta}$.

Secondly, insurance coverage may be unavailable for some losses. For example, \citet{kamiya_risk_2021} found evidence of a potential long-term decrease in sales growth after a data breach. As we are aware, there is no insurance coverage available for such a loss. In addition, as noted in \citet{kamiya_risk_2021}, regulatory fines are common direct costs, but they are often excluded from coverage in cyber insurance policies, as found in \citet{romanosky_content_2019}. If a company wishes to include these losses as impacts on its assets, the indemnity functions $I_{ik}(\cdot)$ for all threat-asset pairs regarding the loss of sales growth and fines should be set to $0$. Nevertheless, the financial implications associated with uninsurable losses can still possibly be mitigated through cybersecurity investments and reserves.


{\subsection{Remarks on cyber risk assessment and cyber capital management revision}
In Section \ref{sec:risk_assess} and this section, a static approach to the cyber risk assessment and cyber capital management is taken. That is, the assessment and capital allocation planning are done at the beginning of a budget period, with the assumption that the risk environment that a company is exposed to does not change significantly during this period. However, upon the arrival of some new knowledge of the risk, the current cyber risk management strategy may no longer be applicable, due to the adaptive nature of cyber risk. Evidence was found in \citet{kamiya_risk_2021} that some attacks may provide additional knowledge about the company's risk exposure, which was not accounted for in the last assessment. For example, a new exploit may reveal the association between a threat and vulnerability pair that was considered non-existent before, and this could result in insufficient funds to mitigate the new risk within the period. Companies may thus wish to regularly review and perform the risk assessment and capital allocation steps described in this paper when substantial changes are detected in the risk environment. For example, if cyber incidents occur during a certain budgeting period, and if the incidents have an impact on the reputation risk in the succeeding period (as suggested by \citet{kamiya_risk_2021}), the impact, such as those direct costs, as well as indirect cost to shareholder value being mentioned in Section \ref{cascade_model_section} earlier, can be incorporated into the budgeting and capital allocation decisions for the next period, through the revised cascade model. In addition, the periodical revision of risk assessment can take any updates in regulations into account. For example, should companies in the US be required to disclose other types of cyber incidents than data breaches in the future, additional compliance costs may be inflicted on companies, and those costs can be included as potential asset losses when the cascade model is revisited during risk assessment.}

\section{Case Study} \label{sec:case}
{\color{black}The case study in the section aims to illustrate the proposed cyber risk assessment and capital allocation framework and derive some economic insights.} 

{\color{black}In practice, a company should map out the cascade structure of its cyber risk according to a specific cybersecurity standard and its own cybersecurity conditions. However, without that confidential information, we can only approximate the cascade model using public records documented in a proprietary dataset compiled by Advisen Ltd. for this case study}. {\color{black} This dataset contains historical records on cyber incidents, and the variables coded for each incident can be put into three categories;}
they are (i) the information of a victim company, (ii) the nature of an incident, and (iii) the consequences, including any associated lawsuits, of the incident.

To gather the data needed as the inputs of the proposed framework, we looked for the threats, vulnerabilities, assets, and losses of incidents from the dataset. Threats are categorized based on the intentions of actors, or lack thereof, that lead to incidents, such as distributed denial-of-service attacks, which cripple the availability of cyber systems, and IT configuration errors, which halt operations unintentionally. Vulnerabilities are parts of a cyber system that threats can directly impact, such as hardware, communication systems, and data systems. Assets recorded in this dataset are items that could lead to direct financial losses if damaged, such as personal financial information (PFI), personally identifiable information (PII), and equipment essential to the continuity of operations. Summary statistics of major categories of threats, vulnerabilities, and assets in this dataset are presented in Table \ref{tab:tva-summary}.

\begin{table}[htbp]
\centering
\resizebox{\columnwidth}{!}{%
{\begin{tabular}{@{}llrrrrr@{}}
\toprule
 &  & All Classes & Class 1 & Class 2 & Class 3 & Other Classes \\ \midrule
\multirow{3}{*}{\begin{tabular}[c]{@{}l@{}}Threat \\ (6 classes)\end{tabular}} & Name & - & Privacy Violation & Data Breach & Extortion/Fraud & - \\
 & Count & 30794 & 23136 & 3700 & 2627 & 1331 \\
 & Percentage & 100.00\% & 75.13\% & 12.02\% & 8.53\% & 4.32\% \\ \midrule
\multirow{3}{*}{\begin{tabular}[c]{@{}l@{}}Vulnerability \\ (5 classes)\end{tabular}} & Name & - & Communication System & Data System & Software & - \\
 & Count & 14470 & 6404 & 4432 & 3134 & 500 \\
 & Percentage & 100.00\% & 44.26\% & 30.63\% & 21.66\% & 3.46\% \\ \midrule
\multirow{3}{*}{\begin{tabular}[c]{@{}l@{}}Asset \\ (12 classes)\end{tabular}} & Name & - & PII & PFI & Business Continuity & - \\
 & Count & 14478 & 10521 & 2332 & 781 & 844 \\
 & Percentage & 100.00\% & 72.67\% & 16.11\% & 5.39\% & 5.83\% \\ \bottomrule
\end{tabular}%
}}
\caption{Summary statistics of major (largest 3) categories of threats, vulnerabilities, and assets}
\label{tab:tva-summary}
\end{table}

Since the proposed framework is for the management of a generic organization's cyber risk, we select one particular victim company from the dataset for this case study. As external observers, we have no access to information about any particular company's cyber system other than the already exposed cybersecurity conditions from the dataset. Hence the company with the most publicly available information is preferred for illustration. Table \ref{tab:inci-num} shows summary statistics of how the observed incidents in this dataset, with the full information on threat, vulnerability, and asset, are distributed by the company and by the industry. Clearly, most of the companies in this dataset have only one incident recorded, which makes it impossible to infer the cascade structure of the cyber risk associated with these companies. Similarly, for most industries, the number of recorded incidents is low, and thus there would be few data points for modeling severity distributions. Therefore, the company with the largest number of observed incidents in the dataset is chosen, and it turns out that it also operates in the industry with the most observations. For confidential courtesy, we shall refer to the victim company as Company X in the sequel.
Note that this {\color{black}preference for a large number of historical cyber incidents} is not required for the practical application of this framework, given that a corporation can conduct a complete assessment of its cyber configurations to learn about its threats, vulnerabilities, and assets.

\begin{table}[htbp]
\centering
\resizebox{\columnwidth}{!}{%
{\begin{tabular}{lrrrrrr}
\hline
 & Minimum & First Quartile & Median & Third Quartile & Maximum & Mean \\ \hline
Incidents per Company & 1 & 1 & 1 & 1 & 236 & 1.8 \\
Incidents per Industry & 4 & 152 & 427 & 1091 & 8759 & 1247 \\ \hline
\end{tabular}%
}}
\caption{Distribution of the number of incidents by company and by industry}
\label{tab:inci-num}
\end{table}

\subsection{Loss tensor}
To assess the cyber risk of Company X, we need to first identify all possible threats, vulnerabilities, and assets, as well as their mapping relationships, for its underlying cyber system. However, the acquired datase{\color{black}t d}oes not contain any sensitive information of a victim company, such as its internal configuration or external exposure. Therefore, as alluded above, we assume that the historical cyber incidents of the victim company have exhaustively covered all possible T$i$, V$j$, and A$k$, for $i=1,2,\dots,l$, $j=1,2,\dots,m$, and $k=1,2,\dots,n$, as well as the mapping matrices $A$ and $B$. As a consequence, we implicitly assume that the victim company could not fully patch the vulnerabilities by only industry-average cybersecurity controls in the past, which is usually the case in practice.

Fo{\color{black}r C}ompany X, there are $l=2$ types of threats, $m=3$ categories of vulnerabilities, and $n=2$ classes of assets, which are summarized in Table \ref{tab:fitting_1}.
\begin{table}[h]
\centering
\begin{tabular}{@{}cr@{}}
\toprule
$i$ & \textbf{Threat}                           \\ \midrule
$1$ & Data Breach                               \\
$2$ & Privacy Violation                         \\ \midrule
$j$ & \textbf{Vulnerability}                    \\ \midrule
$1$ & Communication System                             \\
$2$ & Data System                                      \\
$3$ & Software                                  \\ \midrule
$k$ & \textbf{Asset}                            \\ \midrule
$1$ & Personal Financial Information (PFI)      \\
$2$ & Personally Identifiable Information (PII) \\ \bottomrule
\end{tabular}
\caption{Indexed threats, vulnerabilities, and assets of Company X}
\label{tab:fitting_1}
\end{table}

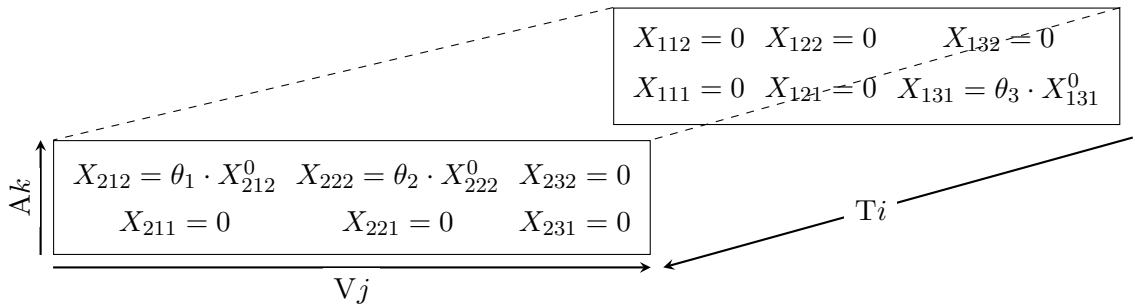
\begin{figure}[htbp!]
   \centering
\begin{tikzpicture}[every node/.style={anchor=north east,fill=white,minimum height=5mm}]
  \matrix (mA_lt) [draw, matrix of math nodes, row sep=1pt,
  column sep=0.2pt, nodes = {anchor=center}]
  {
  X_{112} = 0 & X_{122} = 0 & X_{132} = 0 \\
  X_{111} = 0 & X_{121} = 0 & X_{131} = \theta_3\cdot X_{131}^0  \\
  };
  
  \matrix (mB_lt) [draw,matrix of math nodes, nodes = {anchor=center}] at ($(mA_lt.south west)+(0.5, -0.2)$)
  {
  X_{212} = \theta_1\cdot X_{212}^0 & X_{222} = \theta_2\cdot X_{222}^0 & X_{232} = 0 \\
  X_{211} = 0 & X_{221} = 0 & X_{231} = 0 \\
  };
  
  \draw[thick,-stealth] ([xshift=1ex, yshift=-1ex]mA_lt.south east) -- ([xshift=1ex, yshift=-1ex]mB_lt.south east)
  node[midway,yshift=1ex] {\textsc{T}$i$};
  \draw[thick,-stealth] ([yshift=-1ex]mB_lt.south west) --
  ([yshift=-1ex]mB_lt.south east) node[midway,below] {\textsc{V}$j$};
  \draw[thick,-stealth] ([xshift=-1ex]mB_lt.south west)
  -- ([xshift=-1ex]mB_lt.north west) node[midway,above,rotate=90] {\textsc{A}$k$};
  \draw[dashed](mA_lt.north east)--(mB_lt.north east);
  \draw[dashed](mA_lt.north west)--(mB_lt.north west);
  
\end{tikzpicture}
\caption{Loss tensor of Company X}
\label{fig:fitting_3}
\end{figure}

The mapping matrices of the threats, vulnerabilities, and assets for Company X are given by
\[
A = 
\bordermatrix{%
& \textsc{V1} & \textsc{V2} & \textsc{V3}\cr
\textsc{T1} & 0 & 0 & 1\cr
\textsc{T2} & 1 & 1 & 0
}, \quad
B = 
\bordermatrix{%
& \textsc{A1} & \textsc{A2}\cr
\textsc{V1} & 0 & 1\cr
\textsc{V2} & 0 & 1\cr
\textsc{V3} & 1 & 0
}.
\]
Together with the cybersecurity controls $\bm{\theta} = [\theta_1, \theta_2, \theta_3]$, which shall be determined later by the ex-ante investment allocation $\mathbf{M} = \left[M_1,M_2,M_3\right]$, the cyber loss tensor of Company X is shown in Figure \ref{fig:fitting_3}; the models for the raw random losses $X_{131}^{0}$, $X_{212}^{0}$, and $X_{222}^{0}$ shall be discussed in the following section.

\subsection{Aggregate loss}\label{sec:loss_on_tuple}
To obtain the distributions of the aggregate losse{s $S_{ik}$}, for $i=1,2$, and $k=1,2$, for the purpose of capital allocation, we need to first model the severity for the raw random losses, as well as the frequency for the numbers of cyber incidents. As alluded from the loss tensor of Company X in Figure \ref{fig:fitting_3}, it suffices to model the losses $X_{131}^{0}$, $X_{212}^{0}$, and $X_{222}^{0}$. Moreover, from Figure \ref{fig:fitting_3}, since $Z_{12}=X_{112}+X_{122}+X_{132}=0$ and $Z_{21}=X_{211}+X_{221}+X_{231}=0$, regardless of the realized values of $N_{12}$ and $N_{21}$, we have $S_{12}=S_{21}=0${,} and thu{s i}t suffices to model the total number of cyber incidents $N$, and the numbers of cyber events, $N_{11}$ and $N_{22}$, for the threat-asset pairs (\textsc{T1, A1}) and (\textsc{T2, A2}), of Company X.

Recall that a raw cyber loss should be independent of the company of interest; therefore, all independent cyber losses from the historical incidents, which emerged in the industry where Company X lies in, are used for statistical inference purposes. 
Table \ref{tab:loss_summary} provides their summary statistics from the dataset.


\begin{table}[htbp]
\centering
{\begin{tabular}{lrrr}
\hline
Summary Statistics & $X^0_{131}$ & $X^0_{212}$ & $X^0_{222}$ \\ \hline
Number of Total Observations & 55 & 2253 & 1614 \\
Number of Zero Losses & 17 & 1865 & 1488 \\
Number of Non-Zero Losses & 38 & 388 & 126 \\
Non-Zero Losses Minimum & $6.90 \times 10^1$ & $1.80 \times 10^2$ & $2.31 \times 10^2$ \\
Non-Zero Losses First Quartile & $1.00 \times 10^5$ & $3.99 \times 10^3$ & $3.94 \times 10^3$ \\
Non-Zero Losses Median & $7.53 \times 10^5$ & $2.00 \times 10^4$ & $1.54 \times 10^4$ \\
Non-Zero Losses Third Quartile & $2.54 \times 10^7$ & $2.54 \times 10^5$ & $3.98 \times 10^5$ \\
Non-Zero Losses Maximum & $1.56 \times 10^9$ & $5.05 \times 10^8$ & $1.50 \times 10^7$ \\
Non-Zero Losses Mean & $1.16 \times 10^8$ & $2.67 \times 10^6$ & $8.66 \times 10^5$ \\
\hline
\end{tabular}}
\caption{Summary statistics of industry-wide cyber losses {corresponding to the threats, vulnerabilities, and assets of Company X}}
\label{tab:loss_summary}
\end{table}

From the summary statistics, we can easily infer that zero-inflated models are necessary for fitting these three raw losses well. For the parts of positive loss, several heavy-tailed distributions, such as log-normal, Pareto, and Weibull distributions, are considered. { To choose the most suitable one for this case study, these three probability distributions are fitted to the positive loss data and compared based on their Akaike information criteria (AIC).}

\begin{table}[htbp]
\resizebox{\textwidth}{!}{%
{ \begin{tabular}{@{}lr|rrr|rrr|rrr@{}}
\cmidrule(l){3-11}
\multicolumn{1}{c}{} & \multicolumn{1}{c}{} & \multicolumn{3}{c}{Fitted Log-normal   Distribution} & \multicolumn{3}{c}{Fitted Weibull   Distribution} & \multicolumn{3}{c}{Fitted Pareto   Distribution} \\ \midrule
\multicolumn{1}{c}{} & $q_{ijk}$ & log-mean & log-sd & AIC & shape & scale & AIC & shape & scale & AIC \\ \midrule
$X^0_{131}$ & 0.31 & 12.32 & 3.33 & 16085.59 & 0.30 & 1212057.88 & 16170.87 & 0.32 & 18183.91 & 16163.12 \\
$X^0_{212}$ & 0.83 & 11.95 & 3.09 & 20705.18 & 0.35 & 742665.63 & 20768.82 & 0.31 & 9064.99 & 20816.80 \\
$X^0_{222}$ & 0.92 & 11.43 & 2.94 & 9504.64 & 0.34 & 413025.75 & 9566.94 & 0.34 & 7406.95 & 9527.92 \\ \bottomrule
\end{tabular}%
}}
\caption{Estimated parameters and goodness-of-fit of three positive loss distributions (log-normal, Weibull, and Pareto)}
\label{tab:dist_comparison}
\end{table}

{ Table \ref{tab:dist_comparison} shows the estimated parameters of all three types of fitted probability distributions, where $q_{ijk}$ is the estimated probability mass for zero loss. Their AIC information suggests that log-normal distributions outperform Weibull and Pareto with respect to all three loss random variables $X^0_{131}$, $X^0_{212}$, and $X^0_{222}$. Therefore, in the remaining of this case study, we shall assume that the losses follow log-normal distributions. Note that depending on the loss data and the set of probability distributions being considered, some other options than log-normal could achieve a better fitting result; see, for example, the peak-over-threshold model in \citet{Eling2019}. In this paper, any chosen distributions suffice for the purpose of illustrating the proposed framework.}






Together with the scaling control vector $\bm{\theta} = [\theta_1,\theta_2,\theta_3]$, the cumulative distribution function of the loss random variable $X_{ijk}$ is given by
\begin{equation}
F_{X_{ijk}}\left(x\right)=
\begin{cases}
q_{ijk} & \text{if}\quad x=0\\
q_{ijk}+\left(1-q_{ijk}\right)\Phi\left(\frac{\ln x-\ln\theta_j - \mu_{ijk}}{\sigma_{ijk}}\right)& \text{if}\quad x>0
\end{cases},
\label{eq:impact_distribution}
\end{equation}
for $\left(i,j,k\right)\in\left\{\left(1,3,1\right),\left(2,1,2\right),\left(2,2,2\right)\right\}$, where {$\mu_{ijk}$ and $\sigma_{ijk}$ are the estimated log-mean and log-standard-deviation parameters of the log-normal distribution. These estimated values have been} summarized in Table \ref{tab:dist_comparison}.


The fitted zero-inflated {log-normal} distributions are then discretized for preparing the two convolutions and the mixture in \eqref{eq:Z_ik_dis} and \eqref{eq:L_dis}, in which the probabilities $p_1$ and $p_2$ of the mutually exclusive occurrence of T$1$ and T$2$ for Company X are respectively estimated as $0.015$ and $0.985$.

We fix{ed} one fiscal year for capital allocation purpose{s}. Therefore, the acquired dataset with the historical cyber incidents for Company X from 1997 to 2017 induced 21 observations for each frequency random variable $N$, $N_{11}$, and $N_{22}$. To fit these frequency distributions, we consider two commonly used discrete probability distributions, including Poisson and negative binomial. The fitting results, as shown in Table \ref{tab:freq-dist-compare}, suggest that Poisson distributions are better choices, and the estimated Poisson parameters of the three frequency random variables are $6.48$, $0.1$, and $6.38$, respectively.

\begin{table}[htbp]
\centering
{\begin{tabular}{@{}l|rr|rrr@{}}
\cmidrule(l){2-6}
 & \multicolumn{2}{c}{Fitted Poisson} & \multicolumn{3}{c}{Fitted Negative Binomial} \\ \cmidrule(l){2-6} 
 & mean & AIC & size & mean & AIC \\ \midrule
$N$ & 6.48 & 49.35 & 0.47 & 6.48 & 51.35 \\
$N_{11}$ & 0.1 & 10.09 & 0.04 & 0.09 & 12.09 \\
$N_{22}$ & 6.38 & 47.70 & 0.48 & 6.40 & 49.70 \\ \bottomrule
\end{tabular}%
}
\caption{Estimated parameters and goodness-of-fit of two frequency distributions (Poisson and negative binomial)}
\label{tab:freq-dist-compare}
\end{table}


Finally, as the distribution of the frequency random variables lies in the $\left(a,b,0\right)$ class, we make use of the well-known Panjer recursion to obtain the distributions o{f t}he threat-asset pairing aggregate losses $S_{11}$ and $S_{22}$.

\subsection{Capital {a}llocation}\label{sec:case_allo}

For the cybersecurity investment, we assume that the manager either, invests a fixed amount $\overline{M}_{j}$, or does not allocate any investment, to the cybersecurity control C$j$, in addition to the industry-average level, for the vulnerability V$j$, for $j=1,2,\dots,m$. As a result, if ${M_j}=\overline{M}_{j}$, the losses due to V$j$ are scaled by $\underline{\theta}_{j}\in\left(0,1\right)$, {\it i.e.}, the $j$-th vulnerability is partially, but not fully, patched; if ${M_j}=0$, the V$j$ is only guarded by the industry-standard control, and thus $\theta_j=1$ that Company X shall experience full impact from the raw losses due to V$j$ for its vulnerable cyber system. Therefore, the set of admissible investment allocation $\mathcal{M}=\left\{0,\overline{M}_{1}\right\}\times\left\{0,\overline{M}_{2}\right\}\times\left\{0,\overline{M}_{3}\right\}$, where $\times$ represents the Cartesian product herei{n, a}nd thu{s t}here are {$8$} possible investment allocation{s w}ith the corresponding {$8$} possible cybersecurity control vector{s, w}hich in turn affects the distribution of the impacts via \eqref{eq:impact_distribution}, as well as that of the aggregate losses.

{ For the insurance coverage, we assume that the manager can choose either, not to get any coverages $I_{ik}\left(\cdot\right)\equiv 0$, or to buy a deductible insurance $I_{ik}\left(\cdot\right)=\left(\cdot-d_{ik}\right)_+$, for the aggregate loss $S_{ik}$ of the threat-asset pair (T$i$,A$k$), for $i=1,2,\dots,l$, and $k=1,2,\dots,n$. As a result, if $I_{ik}\left(\mathbf{M}\right)\equiv 0$, $R_{ik}\left(\mathbf{M}\right)=S_{ik}\left(\mathbf{M}\right)$ that the company retains the whole aggregate loss $S_{ik}$; if $I_{ik}\left(\mathbf{M}\right)=\left(S_{ik}\left(\mathbf{M}\right)-d_{ik}\right)_+$, $R_{ik}\left(\mathbf{M}\right)=S_{ik}\left(\mathbf{M}\right)\wedge d_{ik}$ that the company retains the aggregate loss smaller than $d_{ik}$. Therefore, the set of admissible insurance coverage $\mathscr{F}=\left\{0,\left(\cdot-d_{11}\right)_+\right\}\times\left\{0,\left(\cdot-d_{12}\right)_+\right\}\times\left\{0,\left(\cdot-d_{21}\right)_+\right\}\times\left\{0,\left(\cdot-d_{22}\right)_+\right\}$, and thus there are $16$ possible combinations of the insurance coverages.

Hence, the total number of combinations of cybersecurity investments and insurance coverages is $8\times 16=128$. Let $\left(\mathbf{M}_p, \mathbf{I}_p\right)$, for $p=1,2,\dots,128$, be the different pairs of cybersecurity investments and insurance coverages.}

As mentioned in Section \ref{sec:holistic}, the optimal ex-post-loss reserves {$K^*_{ik}\left(\mathbf{M}_p, \mathbf{I}_p\right)$, for $i=1,2$, and $k=1,2$}, shall first be solved for each investmen{t and insurance pair $\left(\mathbf{M}_p, \mathbf{I}_p\right)$, for $p=1,2,\dots, 128$}; and the optimal ex-ante investment $\mathbf{M}^*\in\mathcal{M}$ {and the optimal insurance coverage $\mathbf{I}^*\in\mathscr{F}$} shall then be solved by minimizing the sum of the residual {\color{black}financial implication} of reserve in \eqref{eq:g_r}{,} the {\color{black}financial implication} of cybersecurity investment in \eqref{eq:g_c}{, and the {\color{black}financial implication} of insurance premiums in \eqref{eq:g_i},} among the {$128$} possible investmen{t and insurance combinations}. 

The weighted penalty functions for deviances at individual threat-asset pairs and the aggregate deviance are chosen to be 
$$h_{ik}({R}_{ik}(\mathbf{M})) = \frac{\mathds{1}_{\left\{{R}_{ik}(\mathbf{M}) > \text{VaR}_{0.9}({R}_{ik}(\mathbf{M}))\right\}}}{\mathbb{P}\left({R}_{ik}(\mathbf{M}) > \text{VaR}_{0.9}({R}_{ik}(\mathbf{M}))\right)},$$
for $i = 1,2$ and $k=1,2$, and 
$$h({R}(\mathbf{M})) = \frac{\mathds{1}_{\left\{{R}(\mathbf{M}) > \text{VaR}_{0.9}({R}(\mathbf{M}))\right\}}}{\mathbb{P}\left({R}(\mathbf{M}) > \text{VaR}_{0.9}({R}(\mathbf{M}))\right)},$$
respectively, where $\text{VaR}_{0.9}(\cdot)$ is the Value-at-Risk of a random variable measured at confidence level of $90\%$. Correspondingly, the reserve that minimizes the weighted expected deviance is measured by the Tail Value-of-Risk at the same confidence level.

Moreover, the following are the parameters for cybersecurity investment and control in this case study: $\overline{M}_1=2 \times 10^6$ (\$2 Million), $\overline{M}_2=8 \times 10^6$ (\$8 Million), $\overline{M}_3=10^6$ (\$1 Million), and $\underline{\theta}_1=\underline{\theta}_2=\underline{\theta}_3=0.2$. {Due to the limited availability of open-source data on the costs and effectiveness of cybersecurity solutions, as well as the variations in the functionality and pricing of such products and services by different vendors (see, for example, \citet{vigderman_best_2023}), the cybersecurity investments $\mathbf{M}$ and the mapped control vector $\bm{\theta}$ could only be studied in a ``what-if'' analysis. 
Therefore, hypothetical costs $\overline{\mathbf{M}}$ and effectiveness control vectors $\underline{\bm{\theta}}$ are used in this case study. At the end of this section, we shall provide a sensitivity analysis to show that how changes in cybersecurity costs would affect the company's choice of capital allocation plan.
}

For simplicity, the premium principle $\Pi\left[\cdot\right]$ is assumed to be given by the {\color{black}gross} premium principle{\color{black}, which sets the premium equal to the expected value of loss admitted by the insurer plus incurred expenses. According to \citet{NAIC2022}, the loss ratio of cyber insurers is $66.4$\% on average in 2021. It is reasonable to assume that the expenses account for the remaining $33.6$\%, or approximately $50$\% of the losses, to break even with a combined ratio of $1$. Therefore, we assume that the premium for coverage on each threat-asset pair is $1.5$ times the corresponding expected loss transferred to the insurer, which includes expenses proportional to the claim size.} The deductible $d_{ik}$, for $i=1,2$, and $k=1,2$, is uniformly set to be $\$100,000$ for the insurance coverage of the aggregate loss arising from each threat-asset pair. The rate of opportunity cost $r$ is assumed to be $0.05$ approximating the US one-year treasury rate in 2023. This rate is used to demonstrate the allocation of capital for cyber risk management in the absence of high-yield investment opportunities. Later, an alternative scenario with a larger opportunity cost rate will be presented to highlight the difference in risk management strategies under different capital costs. The relative importance weights $\eta_{j}, {\alpha_{ik}}, \nu_{ik}, \omega_{ik}^\text{I}, \eta, {\alpha}, \nu$, and $\omega^\text{I}$ are set as $1$; that is, all objectives are equally important. 

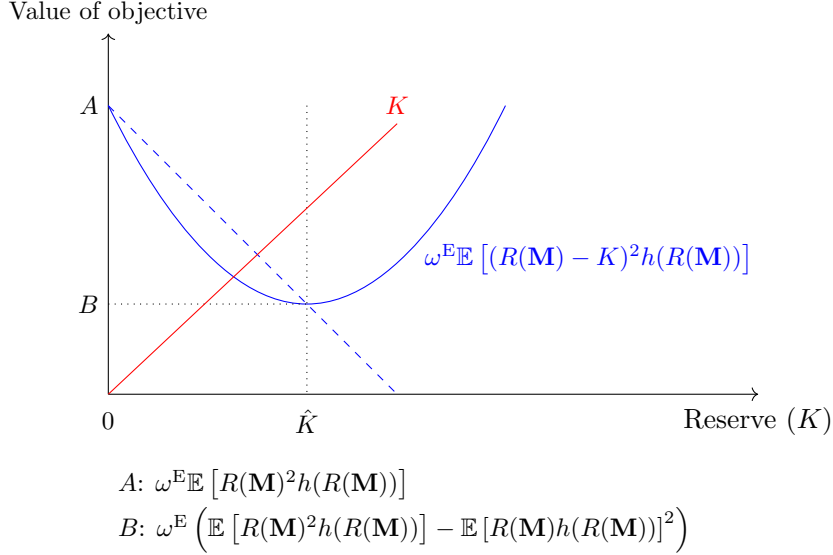
\begin{figure}[ht]
   \centering
   \begin{tikzpicture}[scale=1.2, transform shape, every node/.style={scale=0.8}]

   \draw[->=stealth] (-2.2,0)--(5,0);
   \draw[->=stealth] (-2.2,0)--(-2.2,4) node[above]{\small Value of objective } ;
   \node[] at (-2.2,-0.3) {\small $0$};
   \node[] at (0,-0.3) {\small $\hat{K}$};
   \node[] at (5, -0.3) {Reserve ($K$)};
   \node[left] at (-2.2,3.2) {\small $A$};
   \node[left] at (-2.2,1) {\small $B$};
   \draw[draw = blue](0,1) parabola (2.2,3.2) node[right, blue] at (1.2, 1.5) {\small $\omega^\text{E}\mathbb{E}\left[ (R(\mathbf{M})-K)^2h(R(\mathbf{M}))\right]$};  
   \draw[draw = blue] (0,1) parabola (-2.2,3.2);
   \draw [dotted] (0, 3.2) -- (0, 0);
   \draw [dotted] (-2.2, 1) -- (0, 1);
   \draw [dashed, draw = blue] (-2.2, 3.2) -- (1,0);
   \draw [draw = red] (-2.2, 0) -- (1,3) node[above, red] at (1,3) {\small $K$};
   \node [right] at (-2.2, -1) {\small $A$: $\omega^\text{E}\mathbb{E}\left[{R}(\mathbf{M})^2h({R}(\mathbf{M}))\right]$};
   \node [right] at (-2.2, -1.5) {\small $B$: $\omega^\text{E}\left(\mathbb{E}\left[{R}(\mathbf{M})^2h({R}(\mathbf{M}))\right] - \mathbb{E}\left[{R}(\mathbf{M})h({R}(\mathbf{M}))\right]^2\right)$};
   \end{tikzpicture}
   \caption{Equating average marginal changes of objectives to determine unit-exchange weight}
   \label{fig:normalizer}
\end{figure}

In the following, we explain a practical choice for the unit-exchange weights, $\omega_{ik}^{\text{E}}$ and $\omega^{\text{E}}$. At the optimality, the marginal decrease of the unit-exchanged weighted expected deviance,   $$\omega^\text{E}\mathbb{E}\left[\left({R}(\mathbf{M})-K)^2h({R}(\mathbf{M})\right)\right],$$ coincides with the marginal increase of the opportunity cost of ex-post-loss reserve, $K$, and thus deduces a reasonable unit-exchange weight. However, the optimal ex-post-loss reserve is {\it not known a priori} when the unit-exchange weight is set. Hence, we propose to set the weight by equating the {\it average} marginal decrease of the unit-exchanged weighted expected deviance with the ({\it average}) marginal increase of the opportunity cost of ex-post-loss reserve, $K$. The average is taken at two extreme ex-post-loss reserve allocations; the first one is zero reserve, which is derived when only the objective of opportunity cost is considered; the second one is the optimal standalone reserve $\hat{K}\coloneqq \mathbb{E}\left[ {R}(\mathbf{M})h({R}(\mathbf{M}))\right]$, which is derived when only the objective of unit-exchanged weighted expected deviance is considered. See Figure \ref{fig:normalizer} for its pictorial illustration. Therefore,
\begin{equation*}
\frac{\omega^\text{E}\mathbb{E}\big[({R}(\mathbf{M})- 0)^2h({R}(\mathbf{M}))\big] - 
\omega^\text{E}\mathbb{E}\big[({R}(\mathbf{M})- \hat{K})^2h({R}(\mathbf{M}))\big]}{\hat{K} - 0} = 1,
\end{equation*}
which yields $\omega^\text{E} = \frac{1}{\mathbb{E}\left[ {R}(\mathbf{M})h({R}(\mathbf{M}))\right]}$. Similarly, we propose that, for $i=1,2$ and $k=1,2$, $\omega_{ik}^\text{E} = \frac{1}{\mathbb{E}\left[ {R}_{ik}(\mathbf{M})h_{ik}({R}_{ik}(\mathbf{M}))\right]}$.

\subsubsection*{Without budget constraint}

We shall first assume that Company X has a sufficiently large budget to satisfy the needs for all ex-ante and ex-post-loss {capitals}. Table \ref{tab:num-result} summarizes the findings{, listing the ascendingly sorted, to minimize \eqref{allo_problem1}, selected 8 out of all $128$ feasible combinations of cybersecurity investments and insurance coverages}. Therein, each row shows the corresponding {cybersecurity controls, investment allocations, insurance coverages and premiums, optimal reserves for threat-asset pairs, and the costs of cybersecurity investment, insurance premiums, and reserve.} The highlighted column indicates the case with the optimal {cybersecurity investments and insurance coverages,} which minimize the {total objective function value in \eqref{allo_problem1}}; in this case, the first vulnerability of Company X is additionally patched while the second and the third are left as being guarded only by an industry-average control. Insurance coverage is purchased for losses arising from threat-asset pairs (\textsc{T1, A1}) and (\textsc{T2, A2}).

\begin{table}[ht]
\centering
\resizebox{\textwidth}{!}{%
\begin{tabular}{ll
>{\columncolor[HTML]{FFFE65}}r rrrrrrr}
\hline
\begin{tabular}[c]{@{}l@{}}Cybersecurity   \\ Investment and \\ Insurance Pair \\ Index\end{tabular} & $p$ & 1 & 2 & 3 & 4 & 5 & 6 & 7 & 8 \\ \hline
 & $\theta_1$ & 0.2 & 0.2 & 0.2 & 0.2 & 1 & 1 & 0.2 & 1 \\
 & $\theta_2$ & 1 & 1 & 1 & 1 & 1 & 1 & 0.2 & 0.2 \\
\multirow{-3}{*}{\begin{tabular}[c]{@{}l@{}}Cybersecurity   \\ Controls\end{tabular}} & $\theta_3$ & 0.2 & 0.2 & 0.2 & 0.2 & 1 & 1 & 0.2 & 1 \\
 & $M_1$ & 2 & 2 & 2 & 2 & 0 & 0 & 2 & 0 \\
 & $M_2$ & 0 & 0 & 0 & 0 & 0 & 0 & 8 & 8 \\
\multirow{-3}{*}{\begin{tabular}[c]{@{}l@{}}Investment \\ Breakdown   \\ (millions)\end{tabular}} & $M_3$ & 1 & 1 & 1 & 1 & 0 & 0 & 1 & 0 \\ \hline
 & (\textsc{T1, A1}) & 1 & 1 & 1 & 1 & 1 & 0 & 0 & 0 \\
 & (\textsc{T2, A1}) & 0 & 1 & 0 & 1 & 0 & 0 & 0 & 0 \\
 & (\textsc{T1, A2}) & 0 & 0 & 1 & 1 & 0 & 0 & 0 & 0 \\
\multirow{-4}{*}{\begin{tabular}[c]{@{}l@{}}Insurance on   \\ Threat-asset \\ Pairs \\ ($1$: yes; $0$: no)\end{tabular}} & (\textsc{T2, A2}) & 1 & 1 & 1 & 1 & 1 & 0 & 0 & 1 \\
 & $\pi_{11}\left(\mathbf{M}_p, \mathbf{I}_p\right)$ & 2 & 2 & 2 & 2 & 3.53 & 0 & 0 & 0 \\
 & $\pi_{21}\left(\mathbf{M}_p,   \mathbf{I}_p\right)$ & 0 & 0 & 0 & 0 & 0 & 0 & 0 & 0 \\
 & $\pi_{12}\left(\mathbf{M}_p,   \mathbf{I}_p\right)$ & 0 & 0 & 0 & 0 & 0 & 0 & 0 & 0 \\
\multirow{-4}{*}{\begin{tabular}[c]{@{}l@{}}Premium \\ Breakdown\\ (millions)\end{tabular}} & $\pi_{22}\left(\mathbf{M}_p,   \mathbf{I}_p\right)$ & 4.75 & 4.75 & 4.75 & 4.75 & 7.25 & 0 & 0 & 6.18 \\ \hline
 & $K_{11}^*\left(\mathbf{M}_p, \mathbf{I}_p\right)$ & 0.86 & 0.86 & 0.86 & 0.86 & 0.86 & 31.81 & 17.48 & 32.15 \\
 & $K_{21}^*\left(\mathbf{M}_p,   \mathbf{I}_p\right)$ & 0 & 0 & 0 & 0 & 0 & 0 & 0 & 0 \\
 & $K_{12}^*\left(\mathbf{M}_p,   \mathbf{I}_p\right)$ & 0 & 0 & 0 & 0 & 0 & 0 & 0 & 0 \\
\multirow{-4}{*}{\begin{tabular}[c]{@{}l@{}}Constrained \\ Optimal   \\ Reserves \\ (millions)\end{tabular}} & $K_{22}^*\left(\mathbf{M}_p,   \mathbf{I}_p\right)$ & 4.62 & 4.62 & 4.62 & 4.62 & 4.62 & 28.11 & 12.02 & 4.58 \\ \hline
 & $g_c\left(\mathbf{M}_p\right)$ & 0.3 & 0.3 & 0.3 & 0.3 & 0 & 0 & 1.1 & 0.8 \\
 & $g_i(\mathbf{M}_p, \mathbf{I}_p)$ & 0.67 & 0.67 & 0.67 & 0.67 & 1.08 & 0 & 0 & 0.62 \\
 & $g_r\left(\mathbf{M}_p,   \mathbf{I}_p\right)$ & 0.9 & 0.9 & 0.9 & 0.9 & 0.9 & 757.91 & 773.37 & 964.28 \\
\multirow{-4}{*}{\begin{tabular}[c]{@{}l@{}}Financial \\ Implications \\ (millions)\end{tabular}} & Total & 1.87 & 1.87 & 1.87 & 1.87 & 1.98 & 757.91 & 774.47 & 965.7 \\ \hline
\multicolumn{2}{l}{Total Cost (millions)} & 15.23 & 15.23 & 15.23 & 15.23 & 16.26 & 59.92 & 40.5 & 50.91 \\ \hline
\end{tabular}%
}
\caption{Comparison between different investment and capital allocation strategies \textbf{without} total budget constraint.}
\label{tab:num-result}
\end{table}

Several observations can be made from the results in Table \ref{tab:num-result}. Firstly and intuitively, the results corresponding to $p=1,2,3,4$
show that the threat-asset pairs which generate no loss do not require any ex-post-loss reserves. According to Figure \ref{fig:fitting_3}, (\textsc{T1, A2}) and (\textsc{T2, A1}) are associated with no loss with $S_{12}=S_{21}=0$, and thus it would also be indifferent for Company X to get those covered or not. In practice, due to the possible administrative costs associated with purchasing insurance coverage, it may be to the company's advantage to not get covered for those loss-free threat-asset pairs.

Secondly, if equal importance weights are assigned to the objectives associated with different targets, as specified in this example, it is typically non-optimal for the company to use a less-than-comprehensive strategy. For $p=5$, the company retains the losses up to the deductible, but makes no attempt to reduce the risk through cybersecurity investments. In the case where $p=6$, the company makes all possible cybersecurity investments but buys no insurance. For $p=7$, the company simply retains all the risk without insurance and absorbs the losses only by its reserves. All these three scenarios{\color{black}, especially those two without insurance,} lead t{\color{black}o more substantial financial implications} than the optimal case, in which risk reduction, transfer, and retention are all adopted. At the end of this section, we shall discuss the cases in which the company has strong preferences for some risk management targets over others. Lastly, the case where $p=8$ corresponds to the worst-case scenario among all $128$ possible strategies. Although the company uses various risk management devices, the {\color{black} financial implication} is multiple of that in the optimal case because the capital is allocated poorly to the targets in the cascade model. This example highlights the necessity of a capital allocation framework for corporate cyber risk management, such as the one proposed in this paper.

\subsubsection*{With budget constraint}
\begin{table}[ht]
\centering
\resizebox{\textwidth}{!}{%
\begin{tabular}{llrrr
>{\columncolor[HTML]{FFFE65}}r rrrr}
\hline
\begin{tabular}[c]{@{}l@{}}Cybersecurity   \\ Investment and \\ Insurance Pair \\ Index\end{tabular} & $p$ & 1 & 2 & 3 & 4 & 5 & 6 & 7 & 8 \\ \hline
 & $\theta_1$ & 0.2 & 1 & 0.2 & 1 & 1 & 0.2 & 1 & 1 \\
 & $\theta_2$ & 1 & 1 & 0.2 & 1 & 1 & 0.2 & 1 & 0.2 \\
\multirow{-3}{*}{\begin{tabular}[c]{@{}l@{}}Cybersecurity   \\ Controls\end{tabular}} & $\theta_3$ & 0.2 & 1 & 0.2 & 0.2 & 0.2 & 0.2 & 1 & 1 \\
 & $M_1$ & 2 & 0 & 2 & 0 & 0 & 2 & 0 & 0 \\
 & $M_2$ & 0 & 0 & 8 & 0 & 0 & 8 & 0 & 8 \\
\multirow{-3}{*}{\begin{tabular}[c]{@{}l@{}}Investment \\ Breakdown   \\ (millions)\end{tabular}} & $M_3$ & 1 & 0 & 1 & 1 & 1 & 1 & 0 & 0 \\ \hline
 & (\textsc{T1, A1}) & 1 & 1 & 1 & 1 & 0 & 0 & 0 & 0 \\
 & (\textsc{T2, A1}) & 0 & 1 & 1 & 0 & 0 & 0 & 0 & 1 \\
 & (\textsc{T1, A2}) & 0 & 1 & 1 & 0 & 0 & 0 & 0 & 1 \\
\multirow{-4}{*}{\begin{tabular}[c]{@{}l@{}}Insurance on   \\ Threat-asset \\ Pairs \\ ($1$: yes; $0$: no)\end{tabular}} & (\textsc{T2, A2}) & 1 & 1 & 1 & 0 & 0 & 0 & 0 & 1 \\
 & $\pi_{11}\left(\mathbf{M}_p, \mathbf{I}_p\right)$ & 2 & 3.53 & 2 & 2 & 0 & 0 & 0 & 0 \\
 & $\pi_{21}\left(\mathbf{M}_p,   \mathbf{I}_p\right)$ & 0 & 0 & 0 & 0 & 0 & 0 & 0 & 0 \\
 & $\pi_{12}\left(\mathbf{M}_p,   \mathbf{I}_p\right)$ & 0 & 0 & 0 & 0 & 0 & 0 & 0 & 0 \\
\multirow{-4}{*}{\begin{tabular}[c]{@{}l@{}}Premium \\ Breakdown\\ (millions)\end{tabular}} & $\pi_{22}\left(\mathbf{M}_p,   \mathbf{I}_p\right)$ & 4.75 & 7.25 & 3.67 & 0 & 0 & 0 & 0 & 6.18 \\ \hline
 & $K_{11}^*\left(\mathbf{M}_p, \mathbf{I}_p\right)$ & 0 & 0 & 0 & 0 & 0 & 0 & 0 & 0 \\
 & $K_{21}^*\left(\mathbf{M}_p,   \mathbf{I}_p\right)$ & 0 & 0 & 0 & 0 & 0 & 0 & 0 & 0 \\
 & $K_{12}^*\left(\mathbf{M}_p,   \mathbf{I}_p\right)$ & 0 & 0 & 0 & 0 & 0 & 0 & 0 & 0 \\
\multirow{-4}{*}{\begin{tabular}[c]{@{}l@{}}Constrained \\ Optimal   \\ Reserves \\ (millions)\end{tabular}} & $K_{22}^*\left(\mathbf{M}_p,   \mathbf{I}_p\right)$ & 0 & 0 & 0 & 1 & 3 & 0 & 4 & 0 \\ \hline
 & $g_c\left(\mathbf{M}_p\right)$ & 0.3 & 0 & 1.1 & 0.1 & 0.1 & 1.1 & 0 & 0.8 \\
 & $g_i(\mathbf{M}_p, \mathbf{I}_p)$ & 0.67 & 1.08 & 0.57 & 0.2 & 0 & 0 & 0 & 0.62 \\
 & $g_r\left(\mathbf{M}_p,   \mathbf{I}_p\right)$ & 13.62 & 13.62 & 13.62 & 384.05 & 797.77 & 849.67 & 871.93 & 1074.26 \\
\multirow{-4}{*}{\begin{tabular}[c]{@{}l@{}}Financial \\ Implications\\ (millions)\end{tabular}} & Total & 14.6 & 14.7 & 15.29 & 384.35 & 797.87 & 850.77 & 871.93 & 1075.68 \\ \hline
\multicolumn{2}{l}{Total Cost (millions)} & 9.75 & 10.78 & 16.67 & 4 & 4 & 11 & 4 & 14.18 \\ \hline
\multicolumn{2}{l}{Feasibility} & no & no & no & yes & yes & no & yes & no \\ \hline
\end{tabular}%
}
\caption{Comparison between different investment, { insurance, and reserve} allocation strategies \textbf{with} total budget constraint.}
\label{tab:num-result-budget}
\end{table}

We next consider the scenario in which Company X has a fixed and limited budget for {\color{black} additional spending on cyber risk management. Two reports, \citet{Bernard2020, aiyer2022cybermarket}, indicate companies spend approximately $0.5$\% of their revenues on cybersecurity and the annual growth in security spending is around $14$\% in recent years. Based on this information and the revenue of Company X, we estimate that the additional budget for cyber risk management in the next fiscal year is \$$4$ million. In that case, the unconstrained optimal plan in Table \ref{tab:num-result} at a cost of {\color{black}\$$15.23$} million is no longer unattainable.}

{Ceteris paribus, under the total budget constraint, some results are summarized in Table \ref{tab:num-result-budget}.} Several observations different from the case without the budget constraint are worth noting. One immediate consequence {is that some combinations of cybersecurity investments and insurance coverages cost more than the budgeted amount and become infeasible. In the table, the pairs $\left(\mathbf{M}_p, \mathbf{I}_p\right)$, for $p=1,2,3,6,8$, are infeasible, and out of all $128$ possible combinations, only {\color{black}$24$} are actually feasible, \textit{i.e.}, the company could afford by a {\color{black}\$$4$} million budget in those cases.

Secondly, the optimal solution under the budget constraint can be substantially inferior to that without the constraint. The highlighted column in Table \ref{tab:num-result-budget} represents the viable and optimal capital allocation strategy under th{\color{black}e b}udget constraint. Compared to the optimal solution in Table \ref{tab:num-result}, the cost of implementing the budget-constrained risk management strategy is reduced from {\color{black}\$$15.23$} million to {\color{black}\$$4$} million, but the value of the {\color{black}total financial implication} is increased significantly from {\color{black}$\$1.87$} million to {\color{black}$\$384.35$} million. {\color{black}The increase can solely be attributed to the expansion in reserve-loss mismatch. That is, although the direct costs of cybersecurity, insurance, and reserves are reduced due to the limited budget, the capital available for covering future losses is insufficient and can lead to more severe financial implications.}

Similar to the previous part, which imposes no budget constraint, we report the scenarios in which the company adopts less-than-comprehensive risk management schemes (see $p=2,6,7$), and they lead to infeasible or non-optimal results.}

\afterpage{
\begin{landscape}
\begin{table}[p]
\centering
\resizebox{\columnwidth}{!}{%
\begin{tabular}{llrrrrrrrrrrrr}
\hline
 &  & \multicolumn{2}{c}{Benchmark} & \multicolumn{2}{c}{\begin{tabular}[c]{@{}c@{}}More\\ Affordable \\ Cybersecurity\end{tabular}} & \multicolumn{2}{c}{\begin{tabular}[c]{@{}c@{}}High \\ Opportunity \\ Cost\end{tabular}} & \multicolumn{2}{c}{\begin{tabular}[c]{@{}c@{}}High \\ Loss-Reserve \\ Mismatch Cost\end{tabular}} & \multicolumn{2}{c}{\begin{tabular}[c]{@{}c@{}}High Sensitivity \\ to Cybersecurity\\ Expenses\end{tabular}} & \multicolumn{2}{c}{\begin{tabular}[c]{@{}c@{}}High Sensitivity\\ to  Cyber Insurance\\ Expenses\end{tabular}} \\ \hline
\multicolumn{2}{l}{Budget Constraint} & \multicolumn{1}{c}{\$4 million} & \multicolumn{1}{c}{unlimited} & \multicolumn{1}{c}{\$4 million} & \multicolumn{1}{c}{unlimited} & \multicolumn{1}{c}{\$4 million} & \multicolumn{1}{c}{unlimited} & \multicolumn{1}{c}{\$4 million} & \multicolumn{1}{c}{unlimited} & \multicolumn{1}{c}{\$4 million} & \multicolumn{1}{c}{unlimited} & \multicolumn{1}{c}{\$4 million} & \multicolumn{1}{c}{unlimited} \\ \hline
\multirow{3}{*}{\begin{tabular}[c]{@{}l@{}}Cybersecurity   \\ Controls\end{tabular}} & $\theta_1$ & 1 & 0.2 & 1 & 0.2 & 1 & 0.2 & 1 & 0.2 & 1 & 1 & 1 & 0.2 \\
 & $\theta_2$ & 1 & 1 & 0.2 & 0.2 & 1 & 0.2 & 1 & 1 & 1 & 1 & 1 & 0.2 \\
 & $\theta_3$ & 0.2 & 0.2 & 0.2 & 0.2 & 0.2 & 0.2 & 0.2 & 0.2 & 1 & 1 & 0.2 & 0.2 \\
\multirow{3}{*}{\begin{tabular}[c]{@{}l@{}}Investment \\ Breakdown   \\ (millions)\end{tabular}} & $M_1$ & 0 & 2 & 0 & 2 & 0 & 2 & 0 & 2 & 0 & 0 & 0 & 2 \\
 & $M_2$ & 0 & 0 & 1 & 1 & 0 & 8 & 0 & 0 & 0 & 0 & 0 & 8 \\
 & $M_3$ & 1 & 1 & 1 & 1 & 1 & 1 & 1 & 1 & 0 & 0 & 1 & 1 \\ \hline
\multirow{4}{*}{\begin{tabular}[c]{@{}l@{}}Insurance on \\ Threat-asset \\ Pairs\\ ($1$: yes; $0$: no)\end{tabular}} & (\textsc{T1, A1}) & 1 & 1 & 1 & 1 & 1 & 1 & 1 & 1 & 1 & 1 & 1 & 1 \\
 & (\textsc{T2, A1}) & 0 & 0 & 0 & 0 & 0 & 0 & 0 & 0 & 0 & 0 & 0 & 0 \\
 & (\textsc{T1, A2}) & 0 & 0 & 0 & 0 & 0 & 0 & 0 & 0 & 0 & 0 & 0 & 0 \\
 & (\textsc{T2, A2}) & 0 & 1 & 0 & 1 & 0 & 1 & 0 & 1 & 0 & 1 & 0 & 1 \\
\multirow{4}{*}{\begin{tabular}[c]{@{}l@{}}Premium \\ Breakdown \\ (millions)\end{tabular}} & $\pi_{11}\left(\mathbf{M}_p, \mathbf{I}_p\right)$ & 2 & 2 & 2 & 2 & 2 & 2 & 2 & 2 & 3.53 & 3.53 & 2 & 2 \\
 & $\pi_{21}\left(\mathbf{M}_p,   \mathbf{I}_p\right)$ & 0 & 0 & 0 & 0 & 0 & 0 & 0 & 0 & 0 & 0 & 0 & 0 \\
 & $\pi_{12}\left(\mathbf{M}_p,   \mathbf{I}_p\right)$ & 0 & 0 & 0 & 0 & 0 & 0 & 0 & 0 & 0 & 0 & 0 & 0 \\
 & $\pi_{22}\left(\mathbf{M}_p,   \mathbf{I}_p\right)$ & 0 & 4.75 & 0 & 3.67 & 0 & 3.67 & 0 & 4.75 & 0 & 7.25 & 0 & 3.67 \\ \hline
\multirow{4}{*}{\begin{tabular}[c]{@{}l@{}}Constrained \\ Optimal \\ Reserves \\ (millions)\end{tabular}} & $K_{11}^*\left(\mathbf{M}_p, \mathbf{I}_p\right)$ & 0 & 0.86 & 0 & 0.86 & 0 & 0.8 & 0 & 0.89 & 0 & 0.86 & 0 & 0.86 \\
 & $K_{21}^*\left(\mathbf{M}_p,   \mathbf{I}_p\right)$ & 0 & 0 & 0 & 0 & 0 & 0 & 0 & 0 & 0 & 0 & 0 & 0 \\
 & $K_{12}^*\left(\mathbf{M}_p,   \mathbf{I}_p\right)$ & 0 & 0 & 0 & 0 & 0 & 0 & 0 & 0 & 0 & 0 & 0 & 0 \\
 & $K_{22}^*\left(\mathbf{M}_p,   \mathbf{I}_p\right)$ & 1 & 4.62 & 0 & 4.62 & 1 & 4.33 & 1 & 4.72 & 0.47 & 4.62 & 1 & 4.62 \\ \hline
\multirow{4}{*}{\begin{tabular}[c]{@{}l@{}}Financial\\ Implications\\ (millions)\end{tabular}} & $g_c\left(\mathbf{M}_p\right)$ & 0.1 & 0.3 & 0.2 & 0.4 & 0.4 & 4.4 & 0.1 & 0.3 & 0 & 0 & 0.1 & 1.1 \\
 & $g_i(\mathbf{M}_p, \mathbf{I}_p)$ & 0.2 & 0.67 & 0.2 & 0.57 & 40.36 & 114.57 & 0.2 & 0.67 & 0.35 & 1.08 & 19.98 & 56.72 \\
 & $g_r\left(\mathbf{M}_p,   \mathbf{I}_p\right)$ & 384.05 & 0.9 & 380.98 & 0.9 & 384.35 & 2.49 & 38394.85 & 35.01 & 386.39 & 0.9 & 384.05 & 0.9 \\
 & Total & 384.35 & 1.87 & 381.38 & 1.87 & 425.11 & 121.46 & 38395.15 & 35.98 & 386.74 & 1.98 & 404.13 & 58.72 \\ \hline
\multicolumn{2}{l}{Total Cost (millions)} & 4 & 15.23 & 4 & 15.15 & 4 & 21.8 & 4 & 15.36 & 4 & 16.26 & 4 & 22.15 \\ \hline
\end{tabular}%
}
\caption{Optimal capital allocations in scenarios with different parameter values. }
\label{tab:num-result-sensitivity}
\end{table}
\end{landscape}
}

{\subsubsection*{Varying parameter values}
Provided that the values of importance weights and costs of cybersecurity solutions are hypothetical in this case study due to the unavailability of empirical data, we vary those values in this part to observe how different parameter values will lead to different capital allocation results.

The optimal allocations in Table \ref{tab:num-result} and Table \ref{tab:num-result-budget} are used as benchmarks, to which the optimal allocation strategy in each of the following alternative scenarios is compared.
    
\vspace{-1em}
\paragraph{\textit{More Affordable Cybersecurity}} In the benchmark example, the cost of reducing the impact of Vulnerability \textsc{V2} is \$$8$ million, which is relatively high compared to the loss reduction it can achieve and the costs of cybersecurity solutions for the other two vulnerabilities. To show that a reasonably priced cybersecurity solution for \textsc{V2} is worth adopting, we assume an equivalently effective cybersecurity product that works on \textsc{V2} and costs only \$$1$ million. As a result, in both the budget-constrained and budget-unconstrained scenarios, such a cybersecurity investment leads to optimal allocations, as suggested by the column ``{\color{black}More} Affordable Cybersecurity'' in Table \ref{tab:num-result-sensitivity}. {Note that if the cost of partially patching $\textsc{V2}$ remains at \$$8$ million, but the effectiveness of the control is increased such that the corresponding $\underline{\theta}_2$ is sufficiently small, a similar conclusion can be drawn because the benefit of investment, in terms of loss reduction, exceeds the cost.}
    
\vspace{-1em}
\paragraph{\textit{High Opportunity Costs}}  
    Instead of an opportunity cost rate of $0.05$, we assume that Company X can choose to invest in some other project yielding a return at an annual rate of $0.2$ if the money were not used for cybersecurity, insurance, and reserves. 
    The column ``High Opportunity Costs'' in Table \ref{tab:num-result-sensitivity} corresponds to this scenario. One interesting finding is that when there is no budget constraint, it becomes optimal for Company X to purchase the expensive control for Vulnerability $\textsc{V2}$ in this situation. The additional security expense reduces the amount of capital needed for insurance and reserves, thus lowering the overall financial implication.

\vspace{-1em}
\paragraph{\textit{High Loss-Reserve Mismatching Costs}} There is a mismatch if the allocated reserve differs from the realized loss. The costs of such mismatches relative to other costs in the objective function are captured by weights $\omega_{ik}^\text{I}$ and $\omega^\text{I}$. When the priorities of reducing opportunity costs and reducing mismatches are comparable, as demonstrated in the benchmark example, \textit{i.e.}, $\nu_{ik} = \nu = \omega_{ik}^\text{I} = \omega^\text{I} = 1$, the insufficiency of the reserve are tolerated to some extent for the competing objective of lowering opportunity costs. Such insufficiency is less bearable if the costs of reserve-loss mismatches are high, \textit{e.g.}, $\omega_{ik}^\text{I} = \omega^\text{I} = 100$, whereas all other weights are $1$. This leads to increases in reserves in the budget-unconstrained case to reduce reserve insufficiency. In the budget-constrained scenario, reserved amounts are unchanged due to the lack of funds. 
    
\vspace{-1em}
\paragraph{\textit{High Sensitivity to Cybersecurity Expenses}} The company may prefer one risk management device over another. For example, investing in cybersecurity can be a less favorable option for managerial reasons, which means the company is sensitive to the prices of cybersecurity solutions. Such sensitivity is captured by parameters $\eta_{j}$ and $\eta$. When their values are relatively high compared to other weights, $e.g.$, $\eta_{j} = \eta = 100$ versus all weights being $1$ in the benchmark example, the company would reduce or even avoid cybersecurity investments, as suggested by the column ``High Sensitivity to Cybersecurity Expenses'' in Table \ref{tab:num-result-sensitivity}. 

\vspace{-1em}
\paragraph{\textit{High Sensitivity to Cyber Insurance Premiums}} Similar to the previous scenario, the company could be sensitive to the cost of insurance. To illustrate this, $\alpha_{ik}$ and $\alpha$ are set to be $100$ while other parameters remain at $1$. Compared to the benchmark scenario, more capital is allocated to cybersecurity investments and reserves, and expenses on cyber insurance are {\color{black}reduced}, as the column ``High Sensitivity to Cyber Insurance Expenses'' in Table \ref{tab:num-result-sensitivity} suggests.
}

{\color{black}
\subsubsection*{Practical implications of the case study}
The case study demonstrates how the proposed framework could be adopted for a real-world cyber risk management problem. In addition, a best effort has been made to make model assumptions realistic, which allows for some practical implications to be derived. 

First, when the weights of various objectives introduced in Section \ref{sec:holistic_allo} are comparable, it is in the best interest of companies to adopt a broad range of risk management tools, including reduction, transfer, and retention, such that the lowest total financial implication of cyber risk can be achieved. 
    
Second, the framework provides an optimal amount of spending for cyber risk management, and a constrained budget could lead to reduced immediate costs but increased future financial implications of cyber risks, mainly owing to the enlarged reserve-loss mismatch. Therefore, it is beneficial for companies to spend more on cyber risk management within their capabilities up to their optimum. 
    
Last, the model output is sensitive to various inputs. Companies should conduct thorough external surveys about the price and effectiveness of cybersecurity solutions and cyber insurance products and perform adequate internal evaluations on the weights of various business objectives. The accuracy of these inputs ensures that the capital allocation scheme output by the proposed framework is relevant and meets the risk management goal.
}

\subsection{Results for companies in the same industry}

The case study so far in this section is conducted for a company with a rich public history of cyber incidents for the quality of the analysis. Nevertheless, to demonstrate the generality of the proposed framework, in this subsection, it is applied to all companies in the same industry as Company X, and we obtain the optimal capital allocation strategies for a total of $274$ companies.

\begin{figure}[!htbp]
    \centering
    \begin{subfigure}{0.48\textwidth}
    \centering
\includegraphics[width = \columnwidth]{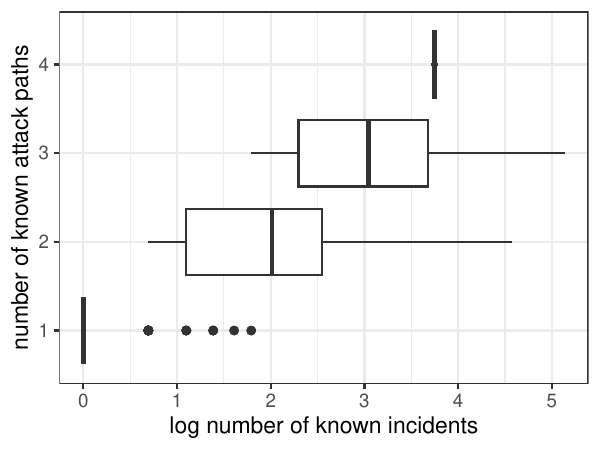}
    \caption{Dependence of the number of attack paths on the number of known incidents}
    \label{fig:inci_num_vs_path_num}
    \end{subfigure}
    \begin{subfigure}{0.48\textwidth}
\centering        \includegraphics[width = \columnwidth]{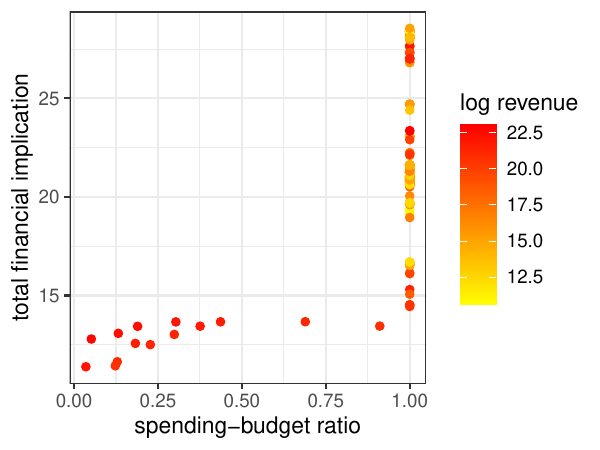}
\caption{Relationship among spending-to-budget ratio, financial implication, and revenue}
\label{fig:spending_implication_revenue}
    \end{subfigure}
    \begin{subfigure}{0.75\textwidth}
        \centering
        \includegraphics[width = \columnwidth]{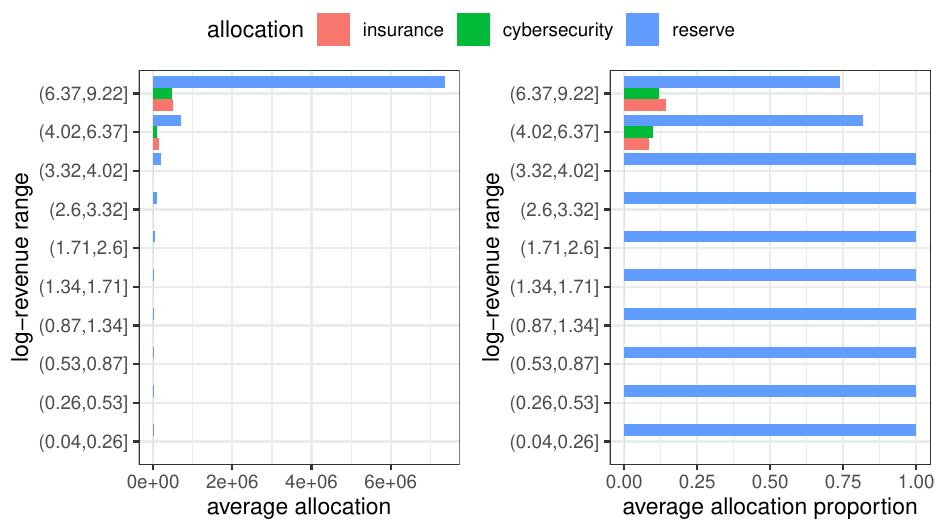}
        \caption{Absolute and relative allocations of capital to insurance, cybersecurity, and reserves for companies with different revenues and with only one historical cyber incident\footnote{To rule out the heterogeneity caused by the number of experienced incidents, only companies that had exactly one incident in the past are used to create these two plots. For companies experiencing multiple incidents, the relationship between their capital allocations and company revenues appears to be similar and thus is omitted for brevity.}}
        \label{fig:allocation_dist}
    \end{subfigure}
    \caption{Cyber risk assessment and capital allocation for other companies in the same industry as Company X}
    \label{fig:empirical}
\end{figure}

Most parameters and model inputs are the same as those of the benchmark case in the last subsection. We recap here for clarity. All relative importance weights are set to $1$, and the unit-exchange weights are determined following the same approach as detailed in Section \ref{sec:case_allo}. Also, being consistent with Section \ref{sec:case_allo}, the insurance premium is set at $1.5$ times the expected claim size, and cybersecurity investment options and their corresponding loss scaling factors are as follows, $\overline{M}_1=2 \times 10^6$ (\$2 Million), $\overline{M}_2=8 \times 10^6$ (\$8 Million), $\overline{M}_3=10^6$ (\$1 Million), and $\underline{\theta}_1=\underline{\theta}_2=\underline{\theta}_3=0.2$.

Some parameters and model inputs are different than the last subsection. There are companies which experienced a fourth type of vulnerability associated with their cyber system hardware, which is not found in Company X's historical incidents. We assume that the investment cost and loss scaling factor corresponding to this vulnerability are $\overline{M}_4=4 \times 10^5$ (\$0.4 Million) and $\underline{\theta}_4=0.2$. A company's budget for additional cyber risk management expenses and the amount of cyber loss to retain are assumed to be proportional to its revenue, with percentages being $0.07\%$ and $0.00175\%$, respectively. These numbers match with Company X's proportions of revenue allocated for the cyber risk management budget and cyber insurance deductible, for fair comparisons.

It is worth noting that, $154$ out of the $274$ companies have only one publicly known historical incident, suggesting that their true risk arrival processes remain largely unknown. Consequentially, the threat-vulnerability-asset mappings and incident frequencies that we estimate for each of these companies are subject to great variability. For this reason, the following discussions focus on general qualitative results and insights among these $274$ companies in the same industry as Company X.



Firstly and most intuitively, assuming that the risk assessment is conducted solely based on publicly available information, the complexity of the mapped attack paths depends on the number of known cyber incidents. Figure \ref{fig:inci_num_vs_path_num} is a box plot and depicts the relationship between the number of unique known attack paths a company is exposed to and its logarithmic number of known incidents\footnote{Log-scaled incident numbers are used for a clearer visual presentation due to the large disparity in number of known incidents.}. It can be observed that a relatively complex risk arrival model with $4$ unique attack paths requires significantly more observed incidents to establish than a simple cascade model with only one attack path. The risk of companies 
with a single attack path may be underestimated due to the inadequate track records. This result highlights the necessity of performing cyber scanning to identify all possible attack paths. Secondly, under the assumptions that budgets are proportional to revenues and that the observed attack paths truly represent the companies' risks, some resourceful companies with low cyber risks may only need to spend a fraction of their budgets to achieve the optimum. Figure \ref{fig:spending_implication_revenue} shows how the ratio between the optimal spending and the budget is related to the size of the company as well as the corresponding financial implication modeled by the objective function in Problem \eqref{allo_problem1}. The budget constraint for small companies is typically binding, thus resulting in a spending-to-budget ratio of $1$, whereas the companies with low spending-to-budget ratios are mostly large in terms of revenue. As pointed out in the previous example about Company X, a binding budget constraint prevents the company from attaining an even lower optimum in the unconstrained case, and therefore, companies with binding budget constraints suffer from higher financial implications caused by cyber risk than those budget-unconstrained companies. Lastly, assuming that budgets are proportional to revenues, small and medium-sized companies may not afford cybersecurity solutions and cyber insurance and will only rely on reserves, whereas large companies often can use more diversified strategies. Figure \ref{fig:allocation_dist} shows the absolute amounts and proportions of the total spending allocated to reserves, cybersecurity solutions, and insurance premiums. When combined with the second finding, an increased budget for cyber risk management allows for a more strategic allocation of capital across different risk management options, ultimately reducing the overall financial implication of cyber risks. 

\section{Concluding Remarks and Future Directions}

In this paper, a cyber cascade model has been proposed for cyber risk assessment and capital management. Classical actuarial and economic models of cyber risks often focus on data trend analysis and do not capture distinct features of cyber risk from other risks. In contrast, engineering literature tends to offer {\color{black}a} microscopic examination of cyber risks, such as attack graphs, which makes it difficult to measure financial impact at a macro level. The model presented in this paper utilizes structural properties of cybersecurity to analyze the cyber risk, through the interactions of threat, vulnerability, control, asset, resulting in financial impact. As an application, we develop a capital allocation scheme in both ex-ante investments and ex-post-loss {capitals}. To deal with the trade-off{s among} cybersecurity investment{s, insurance premiums, and reserves}, as well as to balance the classical competing interests and priorities in enterprise risk management, the optimal ex-ante investment and ex-post-loss {capitals} allocations are solved via the holistic approach. We particularly shed light on the cost-and-benefit analysis between cybersecurity investment, as well as insurance coverage, and the reduction of after-reserve residual risk in this paper. We also apply the proposed framework to one specific company with a large number of historical cyber incidents, as well as to a wide spectrum of companies which yields general insights on cyber risk management.


{Throughout this study, the focus is on cyber losses with quantifiable values. Some losses, such as the loss of reputation, are difficult to estimate owing to the lack of historical data. However, the proposed framework would still be applicable if those losses were quantified using alternative approaches. For example, reputational losses could possibly be estimated using a business analytics approach based on assumptions about the impact of reputation on future cash flows or equity value changes. For other losses, of which the stakes are usually considered too high to analyze from a cost-benefit perspective, such as impacts on national security, a cost-benefit argument can be made nevertheless by stating that benefits are so high that any cost is worthwhile.} 

Evidence for the practicality and usefulness of the proposed framework can be found in some already existing cyber risk management products and services in the industry. Other than the aforementioned tool built by \citet{Thrivaca}, KPMG has also developed a scenario-based cyber risk modeling and quantification model, which has a structure similar to the cascade model used in this study (see \citet{kpmg_cyber_2020}). Compared to the existing solutions in the market, the framework proposed in this paper could provide a scientific foundation, and hence offer more reliable and convincing results in the long run. {\color{black}In addition, we have shown that the cascade structure in the proposed framework resembles elements in CIS Controls, which also overlap regulatory frameworks such as SP 800-53 (see \citet{joint_task_force_interagency_working_group_security_2020}). Because of this, we have the vision that the proposed robust model can potentially be integrated into the regulation.}

Several parts of this paper could {certainly} be further studied. First, we assume that financial losses due to a vulnerability can be linearly reduced by implementing a control. It is possible in practice that the control does not scale down the losses but increases the probability of zero loss {(see, such as, \cite{kamiya_risk_2021})}; in this case, the loss tensor is no longer an element-wise multiplication of the cascade model tensor and the raw loss tensor. But rather, one needs to modify the impact with a point mass at zero loss. 
Second, the case study in this paper is based on the publicly acquirable dataset developed and maintained by Advisen Ltd., which could not reveal sensitive information of victim companies like internal configuration for vulnerabilities, implemented controls, and assets, as well as external exposure for threats. Therefore, we make assumptions for illustrative purposes of the case study. In practice, when implementing the proposed framework, companies could avoid making these assumptions and establish the links among threats, vulnerabilities, and assets based on their own actual cybersecurity conditions (such as by performing cyber scanning). Third, we mentioned that the adopters of this framework may need to frequently revise the risk assessment and capital allocation plan due to substantial changes in the cyber risk environment. Alternatively, a multi-period model can potentially be built to account for those changes and how cyber risk management plans should be updated accordingly. Lastly, given that the purpose of this study is managing the cyber risk of a single organization, from the organization's own perspective, we treat inter-organizational incidents, such as the spread of viruses (contagion), the same way as other incidents without inter-organizational dependence. Future studies may take a multi-agent approach to the cyber risk management for a group of organizations with the loss correlation among them, such as via common vulnerabilities, taken into consideration.



\newpage

\printbibliography

\newpage
\begin{appendices}
\begin{landscape}
\section{\large Mapping Relationship between Control and Asset} \label{append:mapping}
\begin{table}[h]
\footnotesize
\resizebox{0.95\textwidth}{!}{\begin{minipage}{\textwidth}
\begin{tabular}{@{}llllll@{}}
\toprule
\multicolumn{1}{c}{\multirow{2}{*}{\textbf{Control Name}}}                                            & \multicolumn{5}{c}{\textbf{Assets}}                                                                                                                                                              \\
\multicolumn{1}{c}{}                                                                                  & \multicolumn{1}{c}{\textit{Devices}} & \multicolumn{1}{c}{\textit{Applications}} & \multicolumn{1}{c}{\textit{Users}} & \multicolumn{1}{c}{\textit{Network}} & \multicolumn{1}{c}{\textit{Data}} \\ \midrule
1. Inventory and Control of Hardware Assets                                                              & 1                                    &                                           &                                    &                                      &                                   \\
2. Inventory and Control of Software Assets                                                              &                                      & 1                                         &                                    &                                      &                                   \\
3. Continuous Vulnerability Management                                                                   &                                      & 1                                         & 1                                  &                                      &                                   \\
4. Controlled Use of Administrative Privileges                                                           &                                      &                                           & 1                                  &                                      &                                   \\
5. Secure Configuration for Hardware and Software on Mobile   Devices, Laptops, Workstations and Servers &                                      & 1                                         &                                    &                                      &                                   \\
6. Maintenance, Monitoring and Analysis of Audit Logs                                                    &                                      &                                           &                                    & 1                                    &                                   \\
7. Email and Web Browser Protections                                                                     &                                      & 1                                         &                                    & 1                                    &                                   \\
8. Malware Defenses                                                                                      & 1                                    &                                           &                                    & 1                                    &                                   \\
9. Limitation and Control of Network Ports, Protocols and   Services                                     & 1                                    &                                           &                                    &                                      &                                   \\
10. Data Recovery Capabilities                                                                            &                                      &                                           &                                    &                                      & 1                                 \\
11. Secure Configuration for Network Devices, such as   Firewalls, Routers and Switches                   &                                      &                                           &                                    & 1                                    &                                   \\
12. Boundary Defense                                                                                      & 1                                    &                                           & 1                                  & 1                                    &                                   \\
13. Data Protection                                                                                       &                                      &                                           &                                    &                                      & 1                                 \\
14. Controlled Access Based on the Need to Know                                                           &                                      &                                           &                                    & 1                                    & 1                                 \\
15. Wireless Access Control                                                                               & 1                                    &                                           &                                    & 1                                    &                                   \\
16. Account Monitoring and Control                                                                        &                                      &                                           & 1                                  &                                      &                                   \\
17. Implement a Security Awareness and Training Program                                                   &                                      &                                           & 1                                  &                                      &                                   \\
18. Application Software Security                                                                         &                                      & 1                                         &                                    &                                      &                                   \\
19. Incident Response and Management                                                                      & 1                                    & 1                                         &                                    & 1                                    & 1                                 \\
20. Penetration Tests and Red Team Exercises                                                              &                                      &                                           &                                    &                                      &                                   \\ \bottomrule
\end{tabular}
\label{tab:my-table}
\end{minipage}}
\caption{Controls and assets defined in the CIS Controls. Value 1 represents that the control mitigates the vulnerability in the corresponding asset. A blank cell means there is no relationship between the control and the asset.}
\end{table}
\end{landscape}

\begin{landscape}
\section{\large Mapping Relationship between Threat and Control} \label{append:mapping2}
\begin{table}[h]
\footnotesize
\resizebox{0.9\textwidth}{!}{\begin{minipage}{\textwidth}
\begin{tabular}{@{}lcccccccccccccccccccc@{}}
\toprule
\multicolumn{1}{c}{\multirow{2}{*}{\textbf{Threats}}}                 & \multicolumn{17}{c}{\textbf{CIS Controls}}                                                                                                                                                                                         \\
\multicolumn{1}{c}{}                                                  & \textit{1} & \textit{2} & \textit{3} & \textit{4} & \textit{5} & \textit{6} & \textit{7} & \textit{8} & \textit{9} & \textit{10} & \textit{11} & \textit{12} & \textit{13} & \textit{14} & \textit{15} & \textit{16} & \textit{17} & \textit{18} & \textit{19} & \textit{20} \\ \midrule
Tampering (alter physical form or function)                           & 1          &            & 1          &            &            &            &            &            &            &             &             &             &             &             &             &             & & & &            \\
Backdoor (enable remote access)                                       &            & 1          & 1          & 1          & 1          &            &            &            &            &             &             & 1           & 1           &             &             &             &             & & &\\
Use of stolen authentication   credentials                            &            &            &            &            &            &            &            &            &            &             &             &             &             & 1           & 1           &             &             & & &\\
Export data to another site or   system                               &            &            &            &            &            &            &            &            &            & 1           &             &             & 1           &             &             &             & 1           & & &\\
Use of Backdoor or C2 channel                                         &            & 1          & 1          & 1          & 1          &            &            &            &            &             &             & 1           & 1           &             &             &             &             & & &\\
Phishing (or any type of   *ishing)                                   &            & 1          & 1          & 1          & 1          &            &            &            &            &             &             & 1           & 1           &             &             &             &             & & &\\
Command and control (C2)                                              &            & 1          & 1          & 1          & 1          &            &            &            &            &             &             & 1           & 1           &             &             &             &             & & &\\
Downloader (pull updates or   other malware)                          &            & 1          & 1          & 1          & 1          &            &            &            &            &             &             & 1           &             &             &             &             &             & & &\\
Brute force or password guessing attacks                              &            &            &            &            &            &            &            &            &            &             &             &             &             & 1           & 1           &             &             & & &\\
Spyware, keylogger or   form-grabber (capture user input or activity) &            & 1          & 1          & 1          & 1          &            &            &            &            &             &             & 1           &             &             &             &             &             & & &\\
Capture data stored on system   disk                                  &            &            &            &            &            &            &            &            &            &             &             &             &             & 1           & 1           &             & 1           & & &\\
System or network utilities   (e.g., PsTools, Netcat)                 &            & 1          & 1          & 1          & 1          &            &            &            &            &             &             & 1           &             &             &             &             &             & & &\\
Abuse of system access   privileges                                   &            &            &            &            &            &            &            &            &            &             &             & 1           &             & 1           & 1           & 1           &             & & &\\
Ram scraper or memory parser   (capture data from volatile memory)    &            & 1          & 1          & 1          & 1          &            &            &            &            &             &             & 1           &             &             &             &             &             & & &\\
Use of unapproved hardware or   devices                               & 1          &            &            &            &            &            &            &            &            &             &             &             &             &             &             &             & 1           & & &\\
SQL injection                                                         &            &            &            &            &            & 1          &            &            &            &             &             &             &             &             &             &             &             & & &\\
Embezzlement, skimming, and   related fraud                           &            &            &            &            &            &            &            &            &            &             &             &             &             & 1           & 1           &             &             & & &\\
Theft (taking assets without   permission)                            & 1          &            &            &            &            &            &            &            &            &             &             &             &             & 1           &             &             & 1           & & &\\
Bribery or solicitation                                               &            &            &            &            &            &            &            &            &            &             &             & 1           &             & 1           & 1           &             &             & & &\\
Disable or interfere with   security controls                         &            &            & 1          &            &            &            &            &            &            & 1           &             &             &             &             &             &             &             & & &\\
Password dumper (extract   credential hashes)                         &            & 1          & 1          & 1          & 1          &            &            &            &            &             &             & 1           &             &             &             &             &             & & &\\
Misconfiguration                                                      &            &            & 1          &            &            &            &            &            &            & 1           &             &             &             &             &             &             &             & & &\\
Programming error (flaws or   bugs in custom code)                    &            &            &            &            &            & 1          &            &            &            &             &             &             &             &             &             &             &             & & &\\
Misdelivery (direct or deliver   to wrong recipient)                  &            &            &            &            &            & 1          &            &            &            &             &             &             &             &             & 1           &             & 1           & & &\\
Loss or misplacement                                                  & 1          &            & 1          &            &            &            &            &            &            &             &             &             &             &             &             &             & 1           & & &\\ \bottomrule
\end{tabular}
\label{tab:threat-ctrl}
\caption{The mapping between threats and CIS Controls from \citet{SANSInstitute}.}
\end{minipage}}

\end{table}
\end{landscape}

\section{Algorithm for Finding Set {$\mathfrak{I}$}} \label{append:algo}
\SetKw{Break}{break}
\SetKwComment{Comment}{//}{}
\begin{algorithm}[h]
\SetNoFillComment
\caption{Identification Procedure of Set {$\mathfrak{I}$}}\label{alg:set_i}
\KwIn{$\omega_{ik}$ and $\overline{K}_{ik}\left(\mathbf{M},{\mathbf{I}} \right)$, for $i = 1,2,\dots,l$, $k=1,2,\dots,n$; $b\left(\mathbf{M}, {\mathbf{I}}\right)$}
\KwOut{$\mathfrak{I}$}

$\ell \gets l \times n -1$

${\mathfrak{I}}^0 \gets$ list of indices of sorted $\omega_{ik}\overline{K}_{ik}\left(\mathbf{M},{\mathbf{I}}\right)$, such that $\omega_{{\mathfrak{I}}^0[0]}\overline{K}_{{\mathfrak{I}}^0[0]}\left(\mathbf{M},{\mathbf{I}}\right) \leq \omega_{{\mathfrak{I}}^0[1]}\overline{K}_{{\mathfrak{I}}^0[1]}\left(\mathbf{M},{\mathbf{I}}\right) \leq \dots \leq \omega_{{\mathfrak{I}}^0[\ell]}\overline{K}_{{\mathfrak{I}}^0[\ell]}\left(\mathbf{M},{\mathbf{I}}\right)$  

\For{$\iota \gets 0$ to $\ell$}{
${\mathfrak{I}} \gets {\mathfrak{I}}^0[\iota, \dots, \ell]$

\If{$b\left(\mathbf{M}{,\mathbf{I}}\right)+\sum_{\tau=0}^{\ell-\iota} \left(\frac{\omega_{{\mathfrak{I}}[0]}}{\omega_{{\mathfrak{I}}[\tau]}}\overline{K}_{{\mathfrak{I}}[0]}\left(\mathbf{M},{\mathbf{I}}\right)   - \overline{K}_{{\mathfrak{I}}[\tau]}\left(\mathbf{M},{\mathbf{I}}\right)\right) \ge 0$}{
    \Break
}}
\Return{${\mathfrak{I}}$}
\end{algorithm}

\end{appendices}

\end{document}